\documentclass[aps,prx,amsmath,floats,floatfix,twocolumn,
  superscriptaddress,nofootinbib,showpacs]{revtex4-1}

\pdfoutput=1 
\usepackage{diagbox}
\usepackage{mathrsfs}
\usepackage{multirow}
\usepackage{braket}
\usepackage{amsfonts}
\usepackage{amsmath}
\usepackage{amssymb}
\usepackage{amsthm}
\usepackage{bm}
\usepackage{graphicx}
\usepackage{graphics}
\usepackage{latexsym}
\usepackage{rotating}
\usepackage[dvipsnames]{xcolor}
\usepackage{mathrsfs}
\usepackage{microtype}
\usepackage{verbatim}
\usepackage{url}

\usepackage[caption=false]{subfig}
\usepackage[normalem]{ulem}

\usepackage[breaklinks=true]{hyperref}
\hypersetup{
    colorlinks=true,
    linkcolor=NavyBlue,
    filecolor=Magenta,
    urlcolor=NavyBlue,
    citecolor=NavyBlue
}

\usepackage{yfonts}
\usepackage{float}
\usepackage{xspace}
\usepackage{mathrsfs}
\usepackage{siunitx}
\usepackage{array}

\allowdisplaybreaks[4]

\newcommand{\be}{\begin{equation}}
\newcommand{\ee}{\end{equation}}

\newcommand{\ba}{\begin{align}}
\newcommand{\ea}{\end{align}}

\makeatletter
\newcommand*{\rom}[1]{\expandafter\@slowromancap\romannumeral #1@}
\makeatother

\AtBeginDocument{%
    \newwrite\bibnotes
    \def\bibnotesext{Notes.bib}
    \immediate\openout\bibnotes=\jobname\bibnotesext
    \immediate\write\bibnotes{@CONTROL{REVTEX41Control}}
    \immediate\write\bibnotes{@CONTROL{%
    apsrev41Control,author="08",editor="1",pages="1",title="0",year="1"}}
    \if@filesw
    \immediate\write\@auxout{\string\citation{apsrev41Control}}%
    \fi
}

\allowdisplaybreaks

\begin{document}

\title{Black hole ringdown spectroscopy of curvature-coupled scalar and pseudoscalar fields}

\author{Adrian Ka-Wai Chung}
\email{kwc43@cam.ac.uk}
\affiliation{DAMTP, Centre for Mathematical Sciences, University of Cambridge, Wilberforce Road, Cambridge CB3 0WA, United Kingdom}
\affiliation{Illinois Center for Advanced Studies of the Universe \& Department of Physics, University of Illinois Urbana-Champaign, Urbana, Illinois 61801, USA}

\author{Nicol\'as Yunes}
\affiliation{Illinois Center for Advanced Studies of the Universe \& Department of Physics, University of Illinois Urbana-Champaign, Urbana, Illinois 61801, USA}

\date{\today}

\begin{abstract} 

Remnant black holes formed in compact binary coalescences provide a clean setting for probing additional fields coupled to gravity. 
By analyzing the ringdown signals of the loudest LIGO-Virgo-KAGRA binary-black hole mergers, we search for massless curvature-coupled (pseudo)scalar fields arising in extensions to general relativity. 
We find no evidence for such fields and place 90\% credible upper limits on the coupling length scale, $\ell \lesssim (25$-$45)$ km, corresponding to scales significantly below the horizons of the remnant black holes. 
These results include the first gravitational-wave constraint on an axidilaton-type coupling. 
By isolating the ringdown phase and performing extensive robustness checks, our analysis provides a complementary observational probe to existing full-waveform constraints. 
As the gravitational-wave catalog expands, black hole ringdown spectroscopy will enable increasingly sensitive probes of gravity and additional fields using astrophysical black holes.

\end{abstract}

\maketitle

\section{Introduction} 

\label{sec:intro}
Additional fields coupled to gravity may help address several open questions in astronomy and fundamental physics, including the origin of cosmological inflation \cite{Guth:1980zm, Martin:2013tda, Kallosh:2022vha}, the late-time acceleration of the Universe \cite{Saaidi:2012ri, Sa:2021eft, FerreiraJunior:2023qxi}, matter-antimatter asymmetry \cite{Alexander:2004us, Alexander:2009tp}, the nature of dark energy \cite{Shajib:2025tpd}, and possible extensions to general relativity. 
Such fields can modify the spacetime surrounding astrophysical black holes \cite{Cano_Ruiperez_2019, Lam:2025elw, Lam:2025fzi}, leading to deviations from the Kerr description predicted by general relativity. 
Binary black hole mergers observed through gravitational waves therefore provide a promising setting for searching for new fundamental fields \cite{Maselli:2021men, Barsanti:2022vvl, Maselli:2020zgv}, since these modifications can alter black hole dynamics and leave observable imprints on the emitted gravitational-wave signals \cite{Aurrekoetxea:2023jwk, Roy:2025qaa}.

The ringdown phase of binary black hole mergers provides a particularly promising setting for searching for curvature-coupled additional fields. 
During this phase, the newly formed black hole relaxes toward its final stationary state by emitting gravitational waves that can be described as a superposition of exponentially damped oscillations. 
The frequencies and decay times characterizing these quasinormal modes are determined primarily by the spacetime surrounding the remnant black hole and by the properties of the additional fields \cite{Chung:2024vaf}. 
This relative simplicity, together with recent advances in black hole spectroscopy and gravitational-wave data analysis \cite{Berti:2025hly}, has made ringdown observations an increasingly important tool for probing strong-field gravity.

In this work, we search for curvature-coupled (pseudo)scalar fields using recently computed quasinormal-mode spectra of black holes interacting with these fields \cite{Chung:2023wkd, Chung:2024ira, Chung:2024vaf, Chung:2025gyg} together with black hole ringdown signals detected by the LIGO-Virgo-KAGRA detectors. 
We consider three representative couplings: axidilaton, dynamical Chern-Simons, and scalar Gauss-Bonnet couplings. 
These theories have been studied in a variety of contexts connected to inflationary cosmology \cite{Kallosh:2022vha}, Baryon asymmetry \cite{Alexander:2004xd}, and extensions to general relativity. 
They also arise naturally in low-energy effective descriptions of high-energy gravity frameworks, including string theory \cite{Cano_Ruiperez_2019, GROSS19861, Gross:1985rr} and loop quantum gravity \cite{Alexander:2009tp}.

Here, we analyze black hole ringdown signals detected by the LIGO-Virgo-KAGRA detectors to search for curvature-coupled (pseudo)scalar fields. 
Our analysis employs recently computed quasinormal-mode spectra calibrated for remnant black holes with dimensionless spin $a = |\vec{J}|/M^2 \in [0,0.75]$, encompassing the spin range of most binary-black hole merger remnants observed to date. 
We also incorporate the parity-dependent shifts of quasinormal-mode spectra induced by these fields, which have not been explicitly included in previous gravitational-wave searches. 
To assess the reliability of the inferred constraints, we investigate the impact of waveform-modeling and parameter-estimation systematics through a series of robustness tests. 
Finding no evidence for these fields, we place bounds on axidilaton, dynamical Chern-Simons, and scalar Gauss-Bonnet couplings, including the first gravitational-wave constraints on axidilaton-type coupling.

\section{Models of (pseudo)scalar field}\label{sec:theory}

We consider two curvature-coupled (pseudo)scalar fields, $\varphi_1$ and $\varphi_2$, motivated by extensions to general relativity, high-energy gravity frameworks, and cosmology. 
In units with $c=G=1$, which we adopt throughout this work, the fields satisfy
\begin{eqnarray}\label{eq:scalar_fields_eqns}
\Box \varphi_1 = - \ell_1{}^2 \mathscr{G}, 
\qquad
\Box \varphi_2 = - \ell_2{}^2 \mathscr{P}. 
\end{eqnarray}
Here $\Box$ is the wave operator, $\ell_{1,2}$ are coupling length scales with dimensions of length, $\mathscr{G} = R^2 - 4 R_{\alpha \beta} R^{\alpha \beta} + R_{\alpha \beta \gamma \delta} R^{\alpha \beta \gamma \delta}$ is the Gauss-Bonnet invariant, and $\mathscr{P} = R_{\mu \nu \rho \sigma} \tilde{R}^{\mu \nu \rho \sigma}$ is the Pontryagin invariant, where $R_{\alpha \beta \gamma \delta}$, $R_{\alpha \beta}$, and $R$ denote the Riemann tensor, Ricci tensor, and Ricci scalar, respectively, and $\tilde{R}^{\mu \nu \rho \sigma}$ is the dual Riemann tensor.
The choice $(\ell_1,\ell_2) = (\ell,0)$ corresponds to scalar Gauss-Bonnet coupling, while $(\ell_1,\ell_2) = (0,\ell)$ corresponds to dynamical Chern-Simons coupling. 
These couplings arise naturally in low-energy effective descriptions of string theory \cite{Cano_Ruiperez_2019, GROSS19861, Gross:1985rr}, while dynamical Chern-Simons gravity has also been studied in the context of loop quantum gravity \cite{Alexander:2004xd} and baryon asymmetry \cite{Alexander:2004xd}. 
The case $(\ell_1,\ell_2) = (\ell,\ell)$ describes axidilaton coupling, which has been explored in inflationary cosmology \cite{Kallosh:2022vha}. 
Additional details are provided in the Supplementary Materials.

Coupling to these (pseudo)scalar fields deforms the black hole spacetime away from the Kerr solution, thereby shifting the quasinormal-mode spectrum of the remnant black hole. 
Existing gravitational-wave and astrophysical observations already constrain the coupling length scales $\ell_{1,2}$ to be small compared to the black hole mass $M$. 
In this small-coupling regime, the leading-order corrections to the spacetime and quasinormal-mode frequencies scale as $(\ell/M)^4$. 
The quasinormal-mode frequencies can then be expressed as
\begin{equation}\label{eq:QNMFs}
\begin{split}
& \omega^{\rm (P)} (M, a, \ell_1, \ell_2) = \omega^{(0)} (M, a) + \zeta_1 \frac{\Omega_1^{\rm (P)} (a)}{M} + \zeta_2 \frac{\Omega_2^{\rm (P)} (a)}{M},
\end{split}
\end{equation}
where $\omega^{(0)}(M,a)$ denotes the quasinormal-mode frequency of a Kerr black hole with source-frame mass $M$ and dimensionless spin $a$, while $\Omega_{1,2}^{\rm (P)}(a)$ encode the leading-order parity (P)-dependent corrections induced by the couplings. 
We have introduced the dimensionless coupling parameters
\begin{equation}\label{eq:zeta}
\zeta_{1,2} = \left[ \frac{\ell_{1,2} (1+z)}{M_z}\right]^4, 
\end{equation}
where $M_z$ and $z$ are respectively the (redshifted) remnant mass and the redshift to the remnant black hole.
For astrophysical black holes, $\omega^{(0)}(M,a)$ depends primarily on the remnant mass and spin\footnote{Strictly speaking, quasinormal-mode frequencies in general relativity can also depend on electric charge. However, astrophysical black holes are expected to neutralize rapidly through interactions with the surrounding plasma \cite{Blandford:1977ds}.} \cite{Israel:1967za}. 
The functions $\Omega_{1,2}^{\rm (P)}(a)$ for the dominant modes have been computed for spins $a \lesssim 0.75$ \cite{Chung:2024ira, Chung:2024vaf, Chung:2025gyg}. 
For axidilaton coupling, the corresponding corrections are given by the sum of the dynamical Chern-Simons and scalar Gauss-Bonnet contributions, as discussed in the Supplementary Material. 
Throughout this work, we use the fitting expressions for $\Omega_{1,2}^{\rm (P)}(a)$ provided in Refs.~\cite{Chung:2024ira, Chung:2024vaf, Chung:2025gyg}. 
Although parameter estimation occasionally explores spins beyond $a=0.75$, the inferred posteriors are dominated by the region within the calibrated spin range (see Supplementary Material).


\section{Data analysis}\label{sec:DA}

\subsection{Waveform model}

The dependence of the quasinormal-mode spectra on the coupling length scales enables us to build a waveform model that describes the ringdown-phase gravitational waves emitted by black holes coupled to the additional fields \cite{Maenaut:2024oci}, 
\begin{equation}\label{eq:waveform_model}
\begin{split}
h(t) = & \frac{M_z}{d_L} \Theta(t - t_{\rm s}) \sum_{n,l,m,\rm (P)} S_{nlm}(\iota,\phi) \\
& \times \bigg[ \quad A^{+,\rm (P)}_{nlm} e^{-i\omega^{\rm (P)}_{nlm}(t - t_{\rm s})}
+ A^{-,\rm (P)}_{nlm} e^{+i(\omega^{\rm (P)}_{nlm})^{\dagger}(t - t_{\rm s})} \bigg],
\end{split}
\end{equation}
where $ h = h_{+}-ih_{\times}$, with $h_{+, \times}$ being the plus and cross-mode polarization of the gravitational waves respectively, $M_z = M (1 + z)$ is the redshifted mass (i.e.~the mass in the detector frame), $d_L$ is the luminosity distance to the remnant black hole, $\sum_{n,l,m,\rm (P)}$ stands for summation over $n \geq 0, l \geq 2, -l \leq m \leq l$, and parities ((P)), $d_L$ is the luminosity distance to the remnant black hole, and $\Theta(t-t_s)$ is the Heaviside function that activates the waveform at the ringdown start time $t_{\rm s}$. 
The quantity  
$S_{n l m}(\iota, \phi)$ are spin-weighted spheroidal harmonics, which are functions of the inclination angle $\iota$ between the remnant spin axis and the line of sight, and the azimuthal angle $\phi$ in the line of sight and in the source frame. 
The quantities $A^{+, \rm (P)}_{nlm}$ and $A^{-, \rm (P)}_{nlm}$ are complex mode amplitudes, where the  superscript $+$ $(-)$ stands for modes with a positive (negative) real-part frequency.

To simplify the analysis, we adopt two approximations. 
First, we assume $ A^{N, q}_{nlm} = (-1)^l (A^{P, q}_{nlm})^{\dagger} $ \cite{LIGOScientific:2026wpt, Maenaut:2024oci}. 
This is a good approximation for nonprecessing binary-black hole mergers, and is consistent with the events analyzed here (see below), whose inferred remnant spins show no evidence for strong precession \cite{LIGOScientific:2018mvr, Abbott:2020niy, LIGOScientific:2021djp}.
Second, we do not include scalar-led quasinormal modes associated with the additional fields in the present analysis. 
In the small-coupling regime explored here, these modes are expected to enter at higher order in the coupling parameters $\zeta_{1,2}$ and to remain subdominant within the sensitivity of current gravitational-wave observations. 
Our analysis therefore focuses on the leading-order modifications to the gravitational-sector quasinormal-mode spectrum induced by the curvature couplings.

\subsection{Bayesian inference}

The goal of this work is to search for the fields by measuring their coupling length scale $\ell$ using detected ringdown signals via Bayesian inference. 
To this end, We construct the marginalized posterior,
\begin{equation}
p(\ell| d, H, I) = \int d \vec{\theta} ~ p(\ell, \vec{\theta}| d, H, I),
\end{equation}
where $\vec{\theta}$ denotes other source parameters (e.g., $M_z$ and $d_L$), $d$ is the strain data, $H$ is the hypothesis that the signal includes the additional-field effects, and $I$ encodes prior knowledge, such as the quasinormal-mode spectra.
By Bayes' theorem,
\begin{equation}
p(\ell, \vec{\theta}| d, H, I) = \frac{p(d | \ell, \vec{\theta}, H, I)p(\ell, \vec{\theta}| H, I)}{p(d|H, I)},
\end{equation}
where $p(d | \ell, \vec{\theta}, H, I)$ is the likelihood, $p(\ell, \vec{\theta}| H, I)$ is the prior, and $p(d|H, I)$ is the evidence.
If multiple events are analysed, $d = \{d_i|i = 1, 2, 3, \dots\}$, where $d_i$ is the data of the $i$th event, the combined posterior is
\begin{equation}\label{eq:combined_PDF}
p(\ell |{d}, H, I) \propto \prod_{i=1} p(\ell |d_i, H, I),
\end{equation}
since $\ell$ is a coupling constant and the prior of $\ell$ is uniform (see below), which is the same in different events. 
The 90\% credible upper limit $\ell_{90}$, defined by 
\begin{equation}\label{eq:90_CI}
\int_{0}^{\ell_{90}} d \ell ~ p(\ell| {d}, H, I) = 0.9,
\end{equation}
gives us constraints on $\ell$.

\subsection{Parameter estimation}

We analyze publicly available LIGO-Virgo-KAGRA data and focus on events satisfying two criteria; 
(i) a total signal-to-noise ratio $\gtrsim 25$, comparable to that of GW150914, and 
(ii) inclusion in previous ringdown-only tests of general relativity performed by the LIGO-Virgo-KAGRA Collaboration \cite{LIGOScientific:2019fpa, LIGOScientific:2020tif, LIGOScientific:2021sio, LIGOScientific:2026wpt}. 
Applying these criteria, we analyze the ringdown phases of GW150914 \cite{LIGOScientific:2018mvr}, GW190521\_074359 \cite{LIGOScientific:2020ibl}, GW200129\_065458 \cite{LIGOScientific:2021djp}, GW231226\_101520, and GW250114\_082203 \cite{LIGOScientific:2025wao}. 
The inferred progenitor spins of these events show no evidence for strong spin precession \cite{Abbott:2020niy, LIGOScientific:2021djp}, consistent with the symmetry assumption $ A^{-, \rm (P)}_{nlm} = (-1)^l (A^{+, \rm (P)}_{nlm})^{\dagger} $ adopted in Eq.~(\ref{eq:waveform_model}). 

Previous analyses of GW250114\_082203 indicate the presence of an overtone contribution in the ringdown signal \cite{LIGOScientific:2025wao}, and overtone signatures have also been discussed in earlier ringdown studies \cite{Isi:2019aib, Isi:2023nif}. 
In the present analysis, we include only the dominant $nlm = 022$ mode for all events except GW250114\_082203, for which we include both the $022$ and $122$ modes. 
For the overtone modes ($n=1$), we assume vanishing coupling-induced corrections, i.e.,~$\Omega_{1,2}^{\rm (P)}(a) \equiv 0$. 
This treatment is consistent with the fractional-deviation analyses of quasinormal-mode frequencies performed by the LIGO-Virgo-KAGRA Collaboration \cite{LIGOScientific:2026wpt, LIGOScientific:2025wao}. 
As shown in the robustness studies below, including overtone contributions does not significantly alter the inferred constraints.

We analyze the ringdown signals using \texttt{pyRing} \cite{pyRing}, a Python package for time-domain black hole spectroscopy \cite{pyRing_01, pyRing_02, pyRing_03}. 
To reduce contamination from nonlinear dynamics during the early postmerger phase \cite{Cheung:2022rbm, Mitman:2022qdl}, the analysis begins at a time $10M_z$ after the peak of the strain amplitude \cite{pyRing_03}. 
The peak time and remnant mass $M_z$ are estimated \emph{a priori} from full-signal analyses assuming general relativity and are available in the public data releases accompanying each event \cite{LIGOScientific:2020tif, LIGOScientific:2020tif, LIGOScientific:2026wpt}. 
Following standard procedures in black hole spectroscopy \cite{Maenaut:2024oci, LIGOScientific:2020tif, LIGOScientific:2020tif, LIGOScientific:2026wpt}, we fix the sky location and inclination angle to the maximum-likelihood values inferred from full-signal parameter estimation. 
This approximation is justified because these extrinsic parameters are only weakly correlated with the coupling parameters inferred from the ringdown spectrum. 
We prescribe a uniform prior on the following parameters, following~\cite{Maenaut:2024oci}: [$M_z \in [10, 500]\,M_{\odot}$, $a \in [0, 0.93]$, $|A^{\pm,\rm (P)}_{nlm}| \in [0, 50]$, and $\phi^{\pm,\rm (P)}_{nlm} = \arg (A^{\pm,\rm (P)}_{nlm}) \in [0, 2\pi]$]. 
For the luminosity distance $d_L$, we use uniform priors bounded by the 95\% credible interval of the posterior distribution reported by the LIGO-Virgo-KAGRA Collaborations. 
The luminosity distance $d_L$ is converted to redshift assuming a $\Lambda$CDM cosmology, with $H_0=67.9 ~\rm km ~s^{-1} ~ Mpc^{-1}$ the Planck 2015 value of the Hubble constant \cite{Planck:2015fie}. 
The prior on the coupling length scale is taken to be uniform in $\ell \in [0,500]~{\rm km}$. 
To maintain consistency with the perturbative small-coupling treatment, we exclude parameter regions corresponding to $\zeta > 1$. 
We also discard parameter combinations yielding $\mathrm{Im}(\omega)\geq0$, which lie outside the validity regime of the damped-ringdown model adopted here. 
Posterior sampling is performed using 4096 live points, and posterior distributions are reconstructed using kernel density estimation. 
Additional robustness studies examining the dependence on the ringdown start time, astrophysical priors, and small-coupling assumptions are presented in the Supplementary Materials.

\section{Results}\label{sec:results}
\begin{figure}[tp!]
\centering  
\includegraphics[width=\columnwidth]{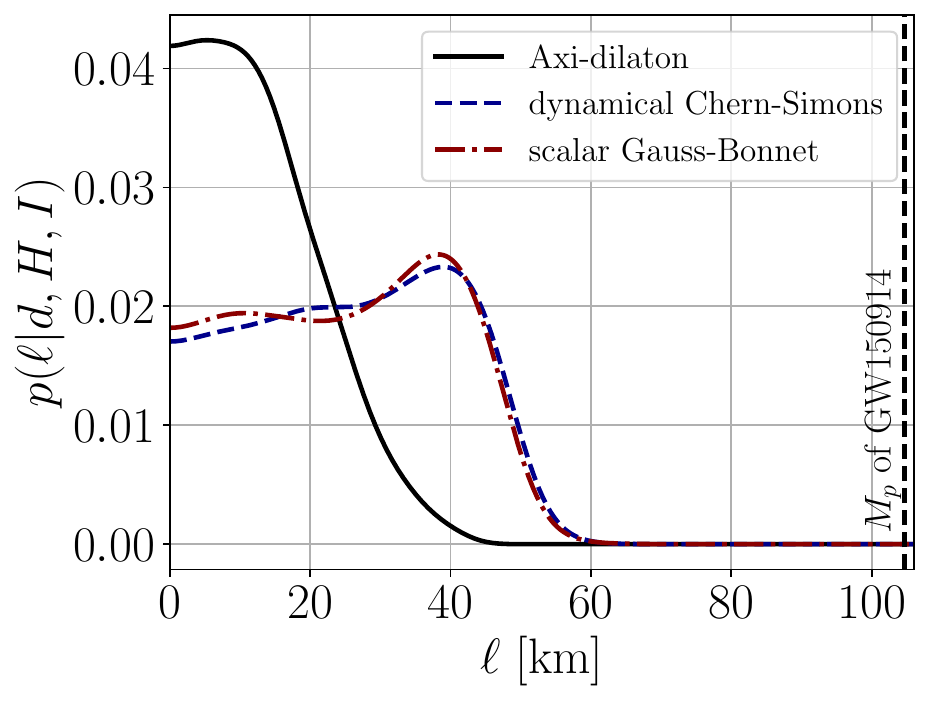}
\caption{Combined marginalized posterior distribution of the coupling length scale, $\ell$ (in km), for scalar fields in axidilaton (solid black), dynamical Chern-Simons (dashed blue) and scalar Gauss-Bonnet (dashed-dotted red) effective theories, obtained by performing Bayesian inference of the ringdown part of the signal whose signal-to-noise ratio $\geq 25$. 
The vertical line indicates the maximum allowed $\ell := M_{\rm map}$ within the small-coupling approximation for the corresponding signal, where $M_{\rm map}$ is the maximal-posterior remnant mass (in km) of the GW150914 event (the least massive of the events for our analysis) in the source frame inferred assuming $\ell=0$. 
}
\label{fig:PDF_ell}
\end{figure}

Figure~\ref{fig:PDF_ell} shows the combined marginalized posterior distribution [c.f. Eq.~\eqref{eq:combined_PDF}] of the coupling length scale, $\ell$ (in km), for (pseudo)scalar fields with the axidilaton (solid black), dynamical Chern-Simons (dashed blue) and scalar Gauss-Bonnet (dashed-dotted red) couplings, obtained by performing Bayesian inference of the ringdown signals of the selected events. 
The vertical line indicates the maximal-posterior remnant mass (in km) of the GW150914 event (the least massive of the events for our analysis) in the source frame inferred assuming $\ell=0$, which represents the longest coupling length scale within which the small-coupling approximation is valid.
We find that all posteriors are consistent with $\ell = 0$, with no evidence for the presence of (pseudo)scalar fields. 
Using Eq.~\eqref{eq:90_CI}, we find that the corresponding 90\% credible upper limits are $\ell \lesssim 25.9$ km, $45.2$ km and $44.4$ km for the axidilaton, dynamical Chern-Simons, and scalar Gauss-Bonnet couplings, respectively. 
Results from individual events are shown in Fig.~1 (in Supplementary Materials) and Table~\ref{Tab:90_CI}. 
Most of the posterior support corresponds to $\zeta \lesssim 10^{-2}$, supporting the leading-order treatment adopted in the waveform model, including the omission of scalar-led quasinormal modes and higher-order spectral corrections. 
A large number of robustness tests are also presented in Supplementary Materials. 

Observe that the posteriors for $\ell$ for fields with axidilaton coupling are the most narrow, with a tighter 90\% credible upper limit, compared to the other quadratic-gravity theories. 
This feature is reasonable because the shift of the quasinormal-mode frequencies with axidilaton coupling is the sum of that in dynamical Chern-Simons and scalar Gauss-Bonnet couplings. 
In the former, axial quasinormal-mode spectra are more significantly modified, while in the latter, stronger modifications appear in the polar spectra. 
Consequently, axidilaton coupling modifies both parity sectors manifestly, making it an easier theory to rule out and leading to the most stringent constraints on $\ell$.

\begin{table*}
\begin{tabular}{c|c|c|c}
\hline
Events  & ~ axidilaton (AD) ~  & ~ Dynamical Chern-Simons (dCS) ~ & ~ Scalar Gauss-Bonnet (sGB) ~ \\ \hline
GW150914 & 40.0 & 53.9 & 57.8\\
GW190521\_074359 & 57.5 & 78.0 & 86.9 \\
GW200129\_065458  & 44.6 & 62.1 & 70.2 \\ 
GW231226\_101520  & 42.3 & 59.9 & 74.5 \\ 
GW250114\_082203  & 35.5 & 48.1 & 50.6 \\ \hline
Combined  & 25.9 & 45.2 &  44.4\\ \hline
\end{tabular}
\caption{The 90\% credible upper limits on $\ell$ (in km), as defined in Eq.~(\ref{eq:90_CI}), for various couplings. 
The constraints are obtained by individually analyzing the ringdown signals selected according to the two criteria stated in the main text, as well as from their combined posterior.}
\label{Tab:90_CI}
\end{table*}

We have verified that our results are robust and consistent with the assumptions underlying our analysis.
First, we have confirmed that all posterior samples satisfy $\zeta \ll 1$, ensuring that the small-coupling approximation remains valid. We have also verified that our constraints on $\ell$ are not artifacts of excluding $\zeta \geq 1$ or $\text{Im} (\omega) > 0$ from the priors. 
The width of the $\ell$ posteriors of the field with dynamical Chern-Simons coupling agrees with previous estimates \cite{Chung:2025gyg}, further supporting the validity of our inference.
We also find no significant degeneracy between $\ell$ and the remnant black hole's mass and spin (i.e., a waveform with nonzero $\ell$ cannot mimic a general-relativity waveform with different $M_z$ and $a$). 
Additionally, we have confirmed that relaxing the approximation $ A^{-, \rm (P)}_{nlm} = (-1)^l (A^{+, \rm (P)}_{nlm})^{\dagger} $ and including an overtone do not significantly affect the constraints.
Finally, we have checked that our results are robust against numerical errors in the quasinormal-mode computations, systematic errors in the quasinormal-mode spectra model for spins $a > 0.75$, and the choice of ringdown start time, provided that merger nonlinearities have sufficiently subsided. 
In particular, the joint posteriors place the vast majority of support within the region $a \lesssim 0.75$, i.e., within the calibrated spin range of the quasinormal-mode spectra used here. 
We also verified that the small-coupling expansion parameter $\epsilon$ remains well below unity for the posterior mass, and that the inferred limits are insensitive to reasonable variations of (i) the treatment of ${\rm Im}(\omega)$ cutoffs, (ii) numerical uncertainties in the parameters of the fitting expression of the quasinormal-mode frequencies, and (iii) the ringdown start time. These checks collectively ensure that the inference remains inside the domain where the underlying quasinormal-mode  modeling is quantitatively reliable. 
Further details are provided in Supplementary Materials.

Our constraints on pseudo(scalar) fields differ from those in~\cite{Silva:2022srr} by a factor of $\sim 4$ because of various differences in analysis methodology. 
First, \cite{Silva:2022srr} included the inspiral, merger and ringdown phases in their analysis, but only considers effects from the pseudo(scalar) fields in the merger and ringdown phase (even though modifications to the merger are not known accurately), while we only consider the ringdown phase.
Their whole-signal analysis recovers a signal-to-noise ratio of $\sim 25$, whereas our ringdown-only analysis yields signal-to-noise ratios $ \lesssim 7$, roughly a factor of 4 smaller.
Second, their analysis uses a model for the quasinormal-mode frequency shift that is accurate only to first order in $a$, which is valid only when $a \lesssim 0.3$; our analysis uses the frequencies in \cite{Chung:2024ira, Chung:2024vaf, Chung:2025gyg} which is accurate up to $a \lesssim 0.75$ and thus applicable to black hole remnants produced in the collision of nonspinning black holes (see also Supplementary Materials).   
Third, their analysis includes only the less-damped parity sector of each field, whereas we incorporate shifts in both axial and polar quasinormal modes. 
Therefore, although the reported constraints in~\cite{Silva:2022srr} are claimed to be more stringent (due to the use of a larger signal-to-noise ratio), the results presented here are more robust. 

\section{Concluding remarks}

Our analysis constitutes a ringdown-based search for scalar fields that incorporates parity-dependent quasinormal-mode spectra calibrated for moderately spinning black holes, explicitly accounting for the breaking of isospectrality. 
We obtain the first gravitational-wave constraint on axidilaton-type coupling, together with bounds on dynamical Chern-Simons and scalar Gauss-Bonnet couplings from the ringdown components of the analysed signals. 
By isolating the ringdown phase, our approach reduces dependence on modeling assumptions spanning multiple dynamical regimes. 
A quantitative comparison with previous analyses is provided in Supplementary Material.

Our results demonstrate that black hole ringdown observations alone can already provide meaningful constraints on additional degrees of freedom beyond general relativity at horizon scales. 
In particular, the absence of detectable parity-dependent effects in the quasinormal-mode spectrum places bounds on curvature-coupled scalar fields that modify the strong-field dynamics of gravity. 
These constraints are complementary to those obtained from full-waveform analyses and other astrophysical probes.

Looking ahead, the sensitivity of ringdown searches will improve as the number of detected events increases. 
Current observations already constrain coupling length scales to values significantly below the horizon size of stellar-mass black holes, and combining multiple events will further tighten these bounds. 
Future developments, including improved modeling of parity-dependent quasinormal-mode spectra and extensions to inspiral-merger-ringdown analyses, will further expand the reach of black hole ringdown spectroscopy as a probe of gravity and additional fields.

\section*{acknowledgements}
The authors acknowledge the support from the Simons Foundation through Award No. 896696, the NSF through Grant No. PHY-2207650 and NASA through Grant No.~80NSSC22K0806. 
A. K. W. C also acknowledges the Herchel Smith Fellowship at the University of Cambridge for support of this work. 
The authors thank Emanuele Berti, Mark Cheung, Carl-Johan Haster, Kelvin K.H. Lam, Tjonnie G.F. Li, Samson H. W. Leong, Hector O Silva, Isaac C.F. Wong, and Sophia Yi for insightful discussion. 
A. K. W. C would like to thank Simon Maenaut and Gregorio Carullo for advise on preparing and sharing the configuration files for \texttt{pyRing} analyses. 
The calculations and results reported in this Letter were produced using the computational resources of the Illinois Campus Cluster, a computing resource that is operated by the Illinois Campus Cluster Program (ICCP) in conjunction with National Center for Supercomputing Applications (NCSA), and is supported by funds from the University of Illinois at Urbana-Champaign, and used Delta at NCSA through allocation PHY240142 from the Advanced Cyberinfrastructure Coordination Ecosystem: Services $\&$ Support (ACCESS) program, which is supported by National Science Foundation Grants No. 2138259, No. 2138286, No. 2138307, No. 2137603, and No. 2138296.
The author would like to especially thank the investors of the IlliniComputes initiatives and GravityTheory computational nodes for permitting the authors to execute runs related to this work using the relevant computational resources. 
\texttt{astropy}~\citep{astropy2022}, \texttt{matplotlib}~\citep{hunter2007matplotlib}, \texttt{numpy}~\citep{harris2020numpy}, \texttt{scipy}~\citep{virtanen2020scipy}, \texttt{pyRing}~\citep{pyRing}

\section*{Data availability} The gravitational-wave strain data analyzed in this work are publicly available through the Gravitational Wave Open Science Center (GWOSC) at \url{https://gwosc.org/eventapi/}. 
The time-domain inference package \texttt{pyRing}, used in this work to search for direct waves, is publicly available at \url{https://git.ligo.org/lscsoft/pyring}.

\bibliography{ref}

\newpage

\appendix

\section{Event-by-event results} \label{sec:Results_event_by_event}

\begin{figure*}[htp!]
\centering  
\subfloat{\includegraphics[width=0.25\textwidth]{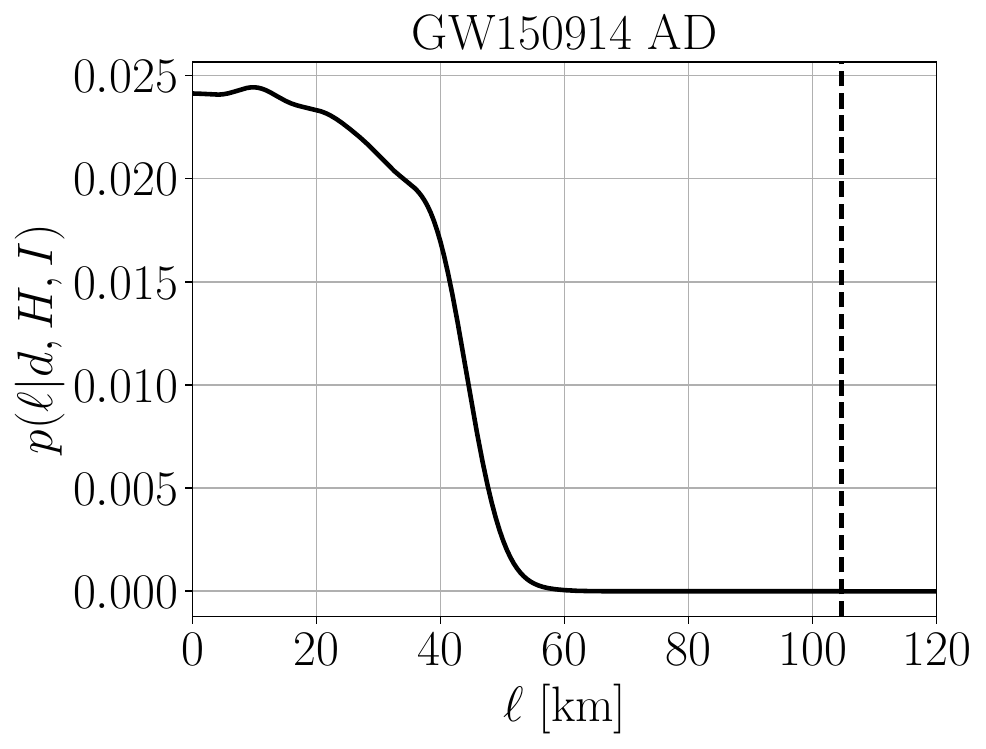}}
\subfloat{\includegraphics[width=0.25\textwidth]{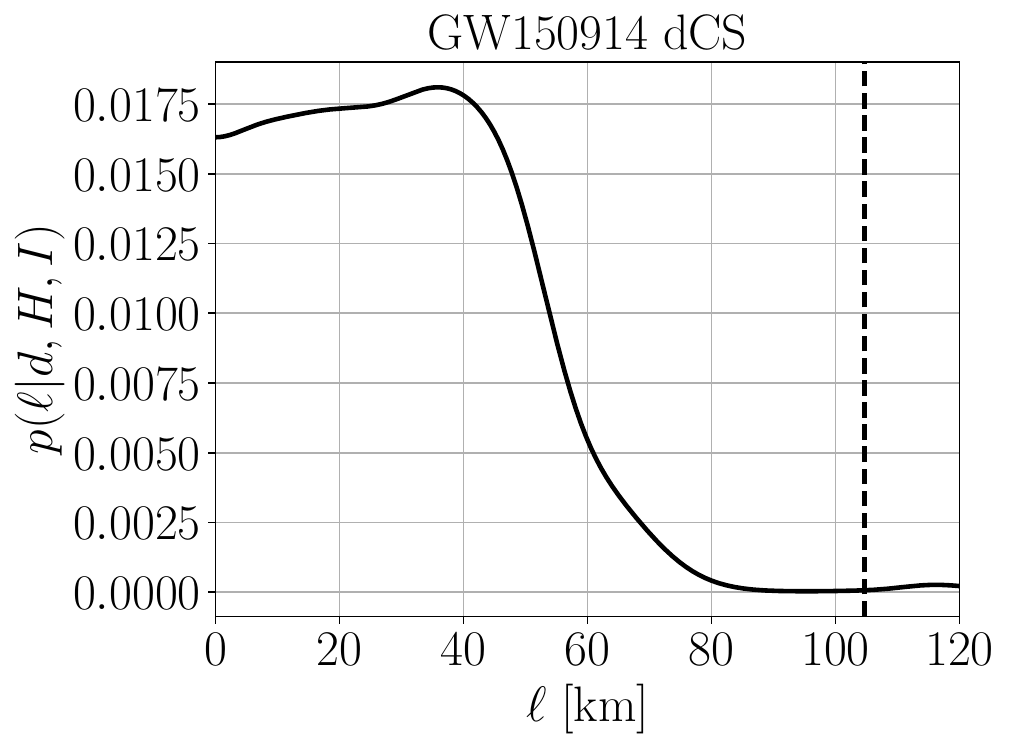}}
\subfloat{\includegraphics[width=0.25\textwidth]{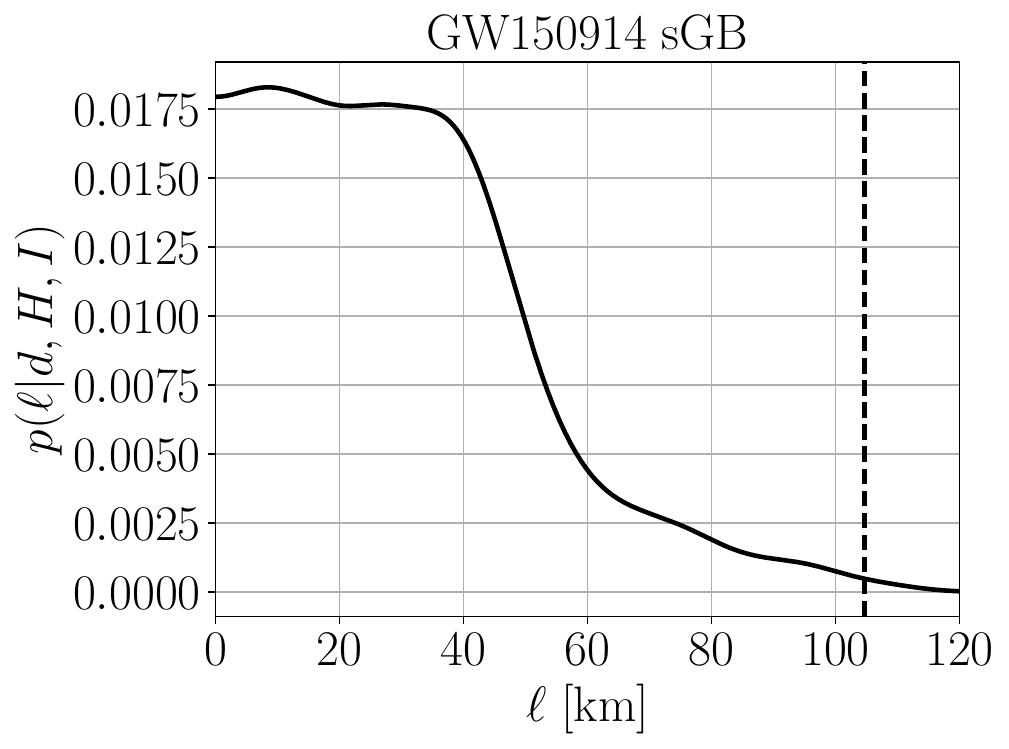}}
\quad
\subfloat{\includegraphics[width=0.25\textwidth]{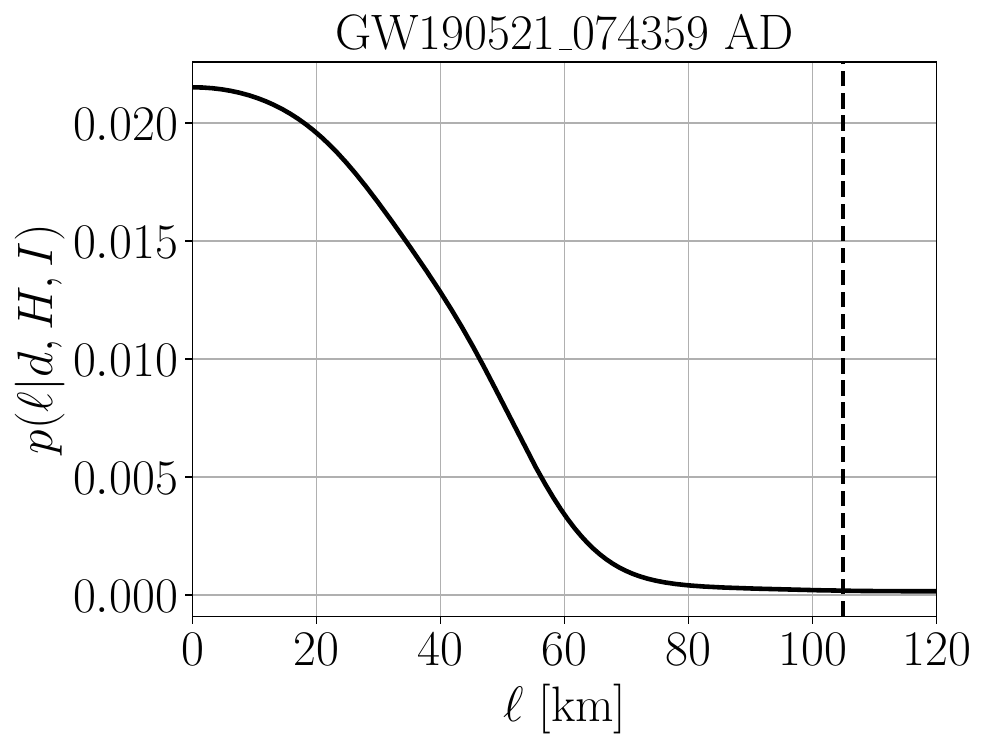}}
\subfloat{\includegraphics[width=0.25\textwidth]{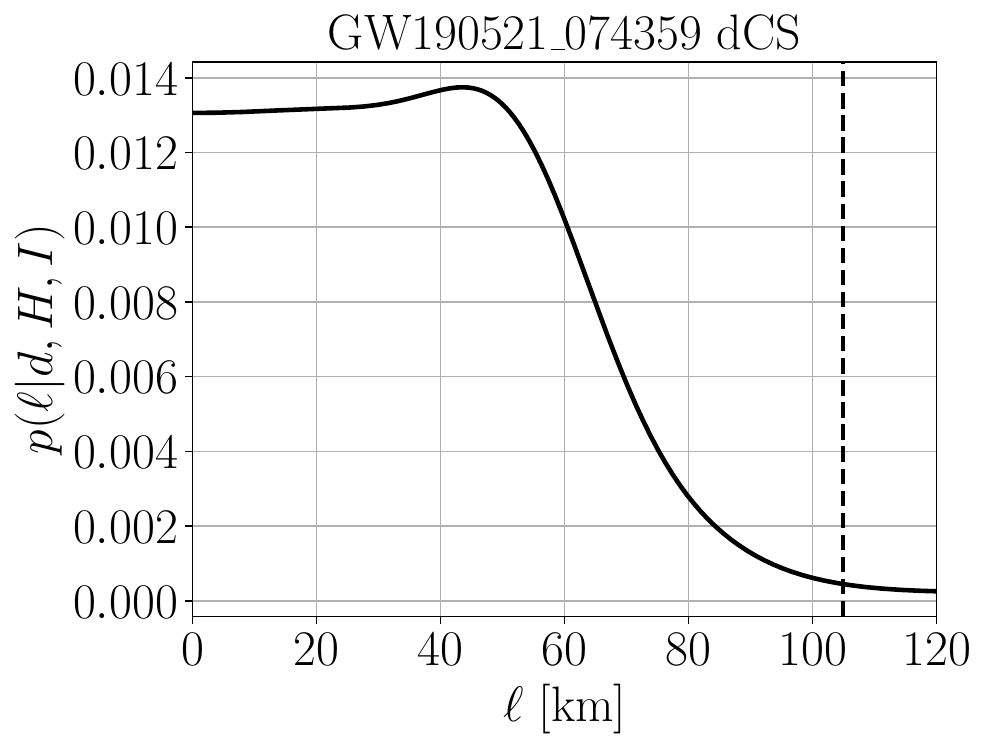}}
\subfloat{\includegraphics[width=0.25\textwidth]{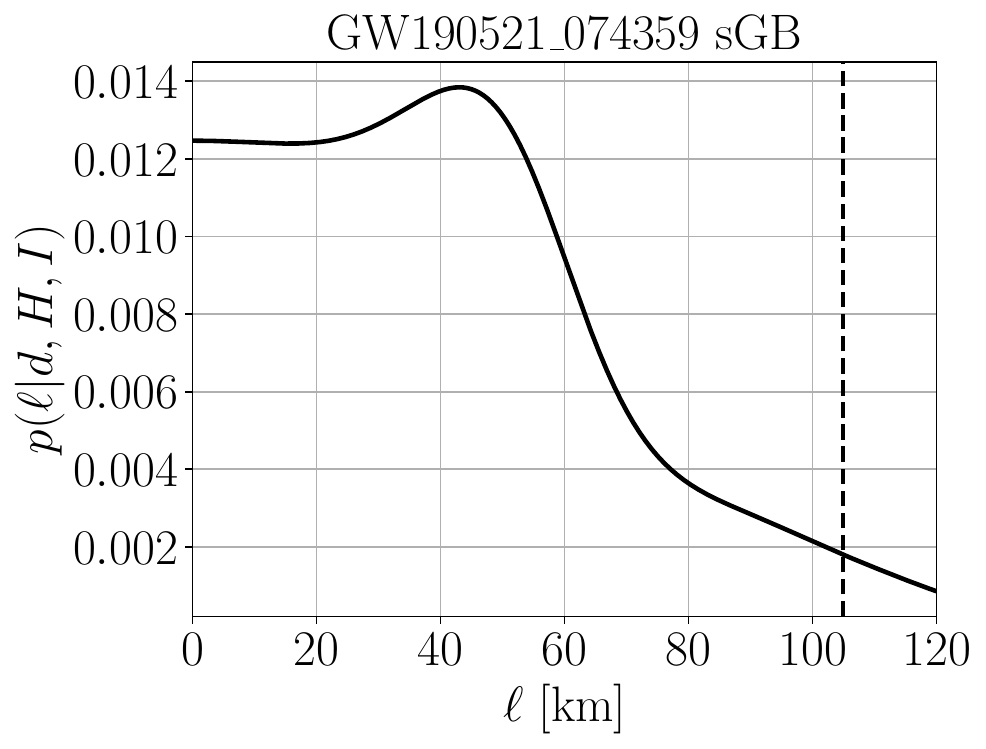}}
\quad
\subfloat{\includegraphics[width=0.25\textwidth]{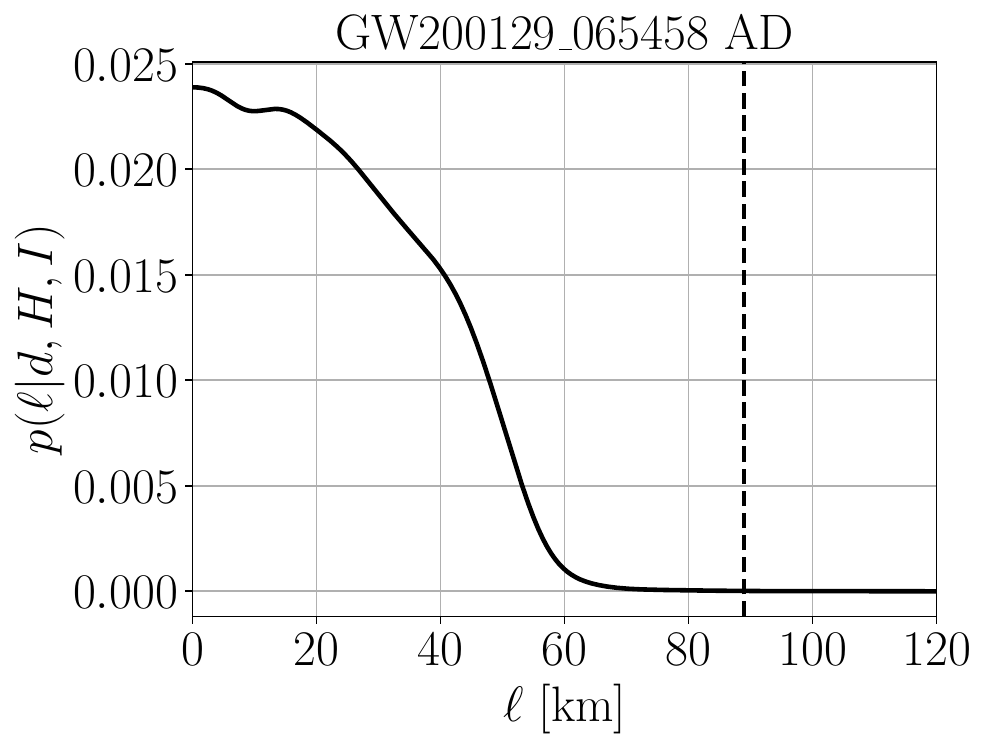}}
\subfloat{\includegraphics[width=0.25\textwidth]{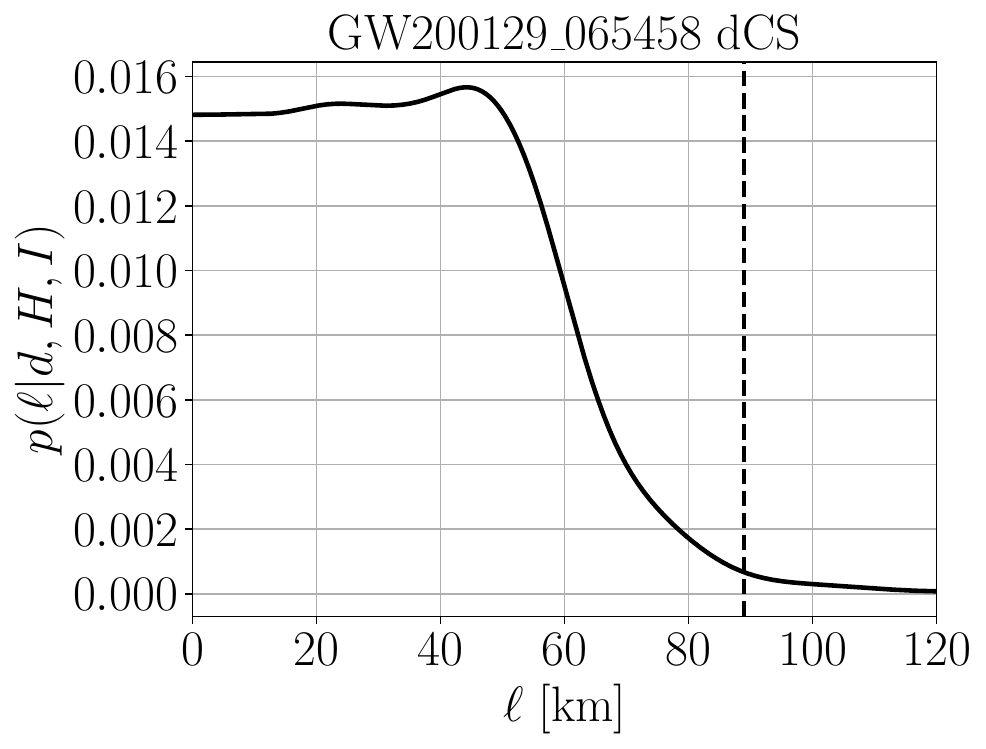}}
\subfloat{\includegraphics[width=0.25\textwidth]{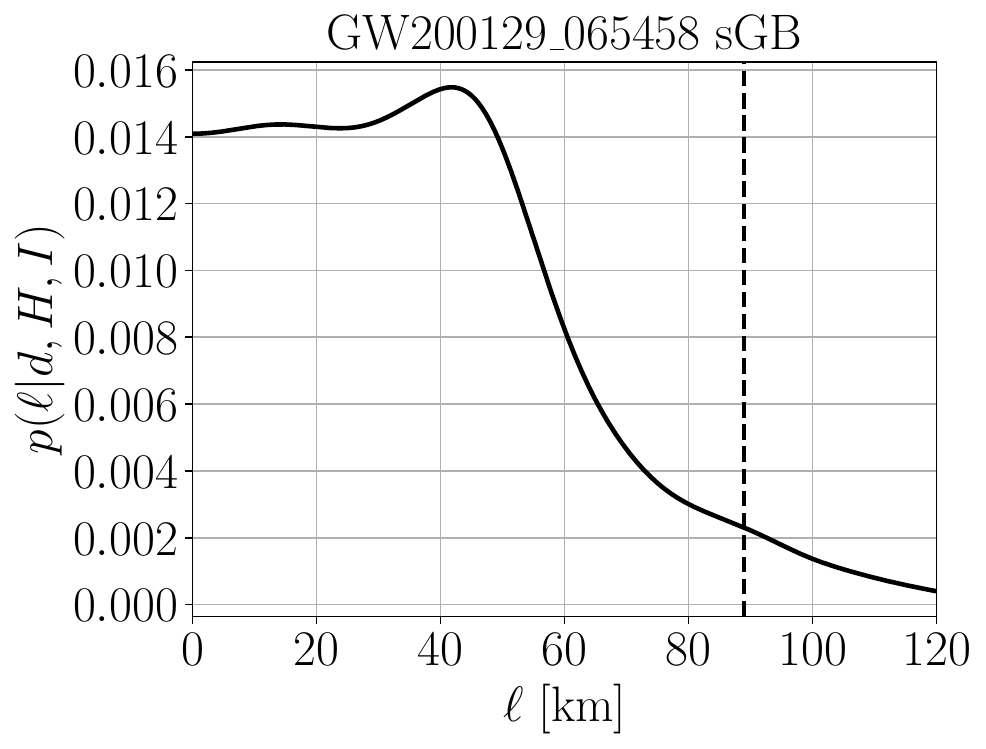}}
\quad
\subfloat{\includegraphics[width=0.25\textwidth]{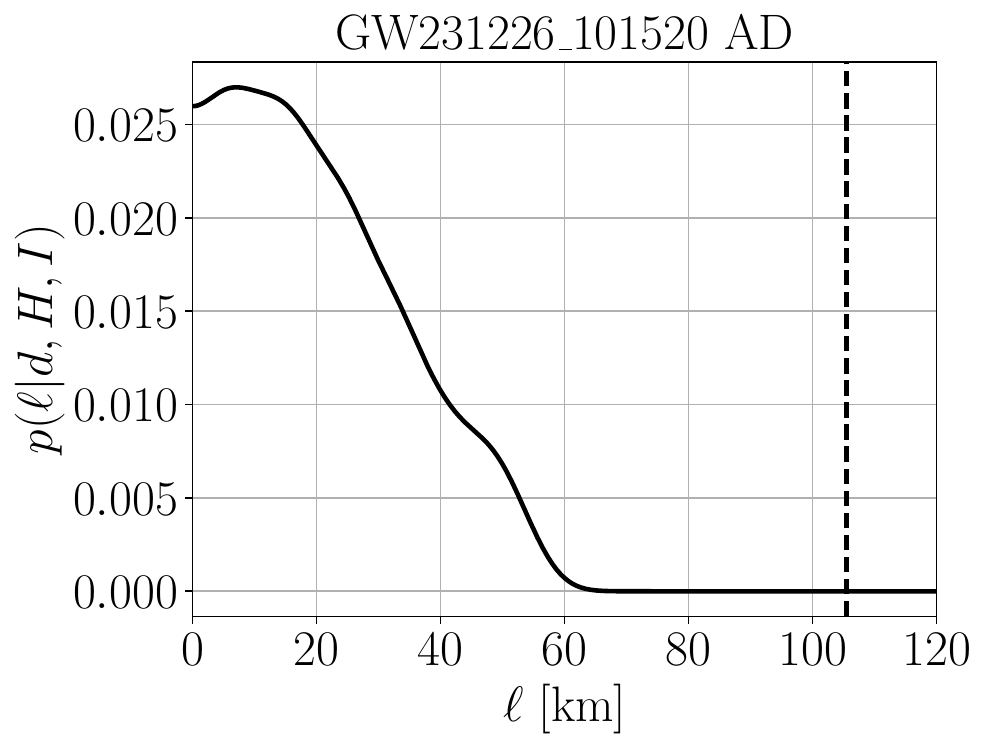}}
\subfloat{\includegraphics[width=0.25\textwidth]{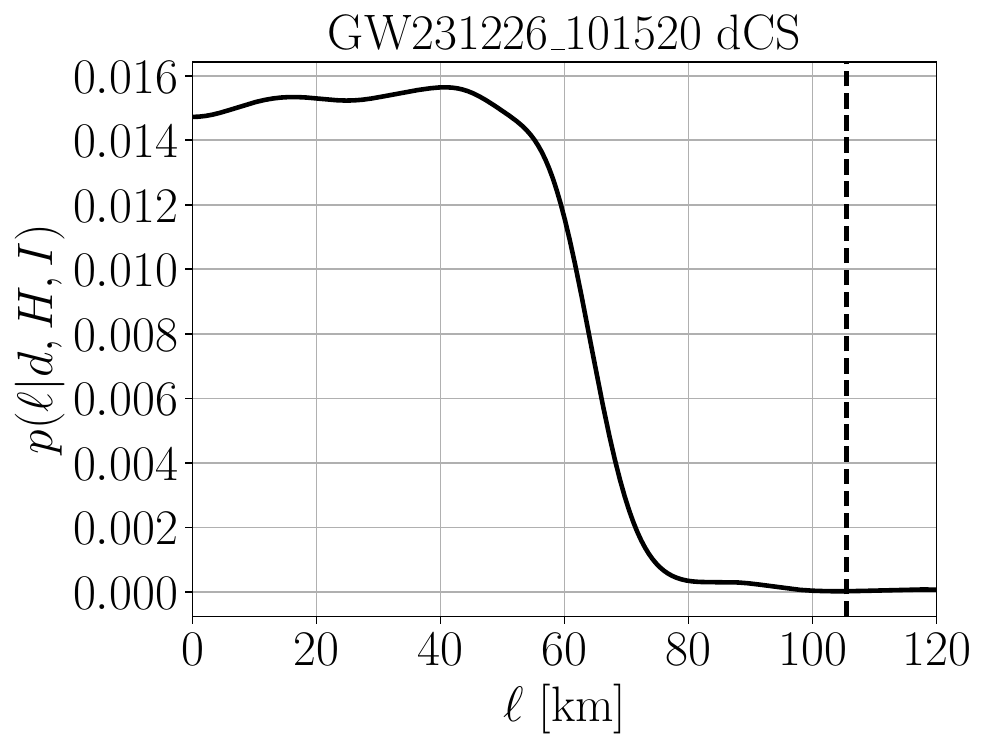}}
\subfloat{\includegraphics[width=0.25\textwidth]{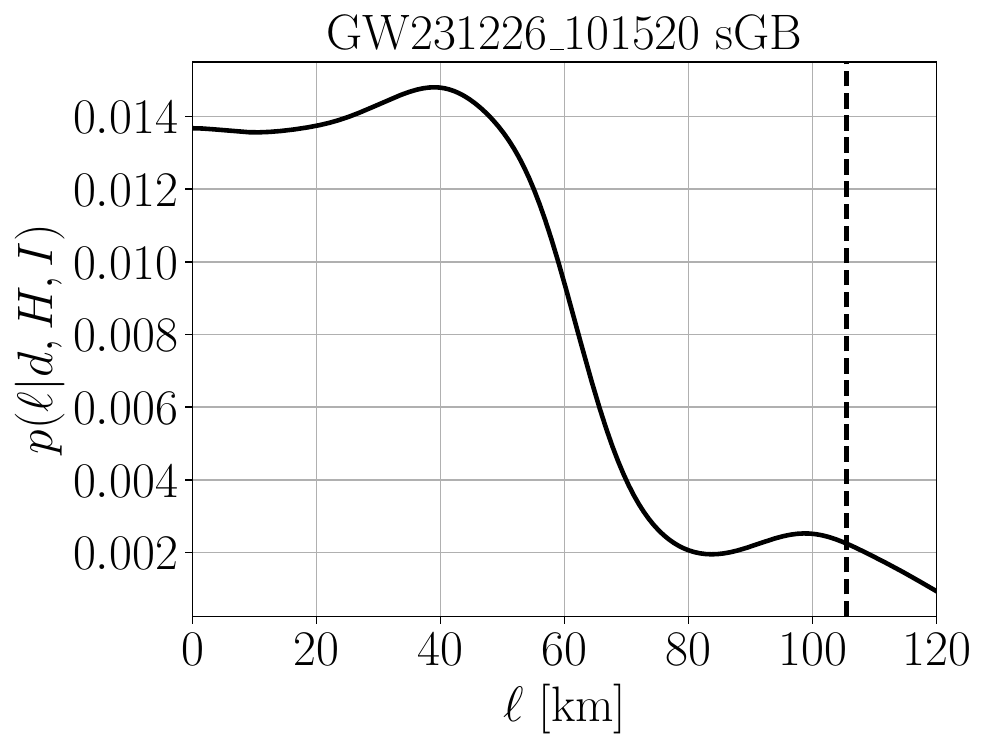}}
\quad
\subfloat{\includegraphics[width=0.25\textwidth]{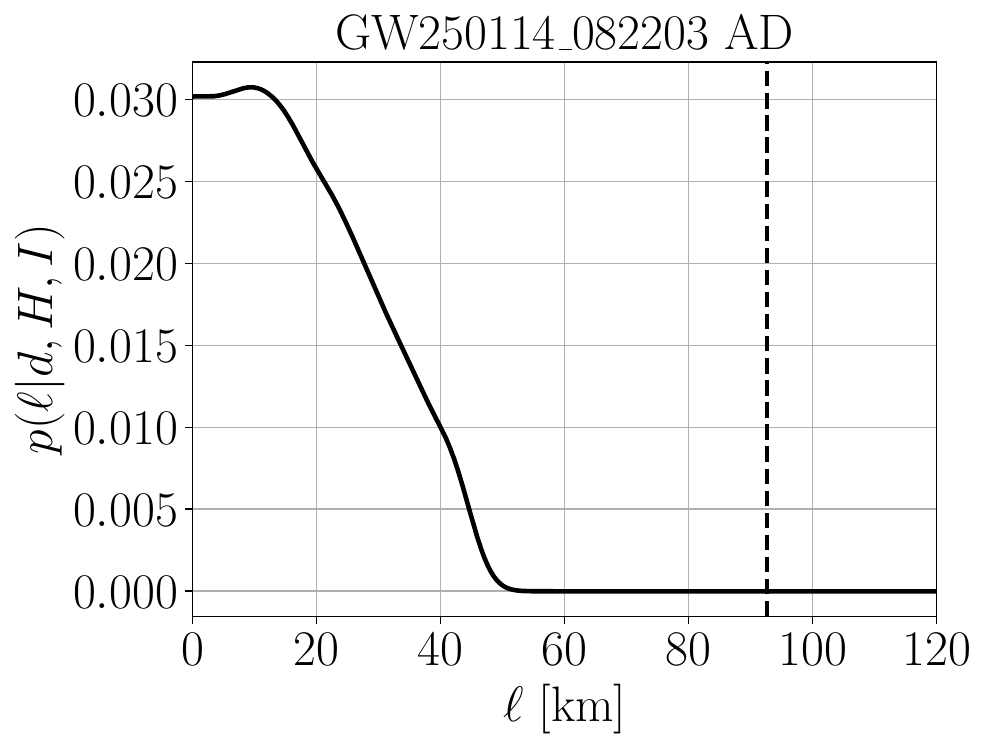}}
\subfloat{\includegraphics[width=0.25\textwidth]{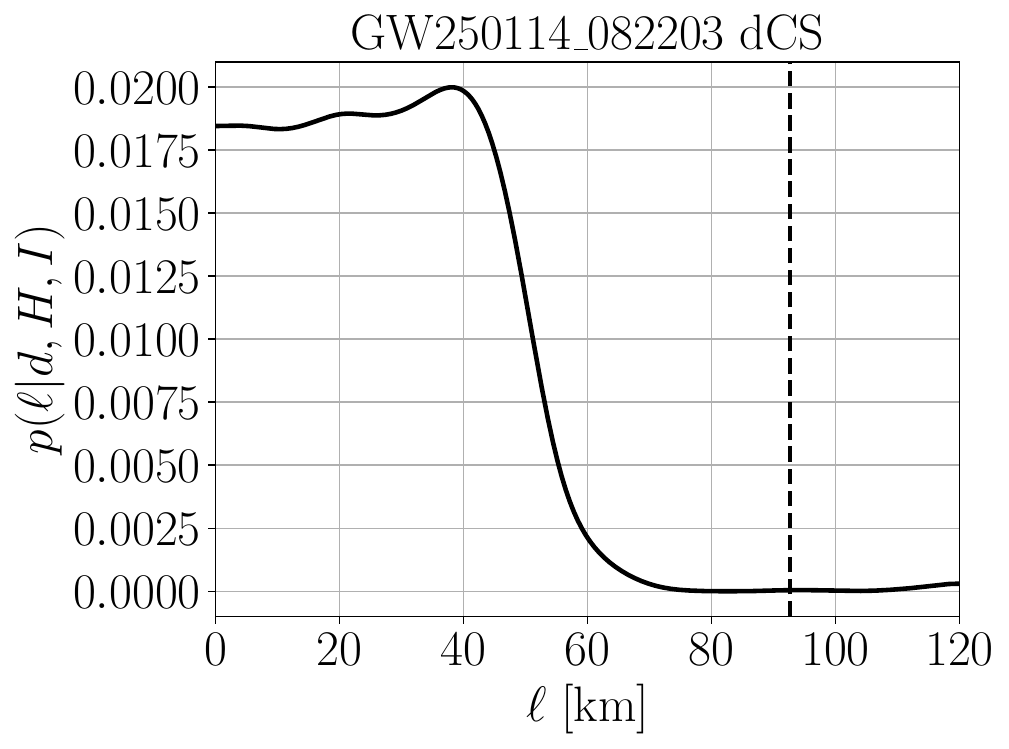}}
\subfloat{\includegraphics[width=0.25\textwidth]{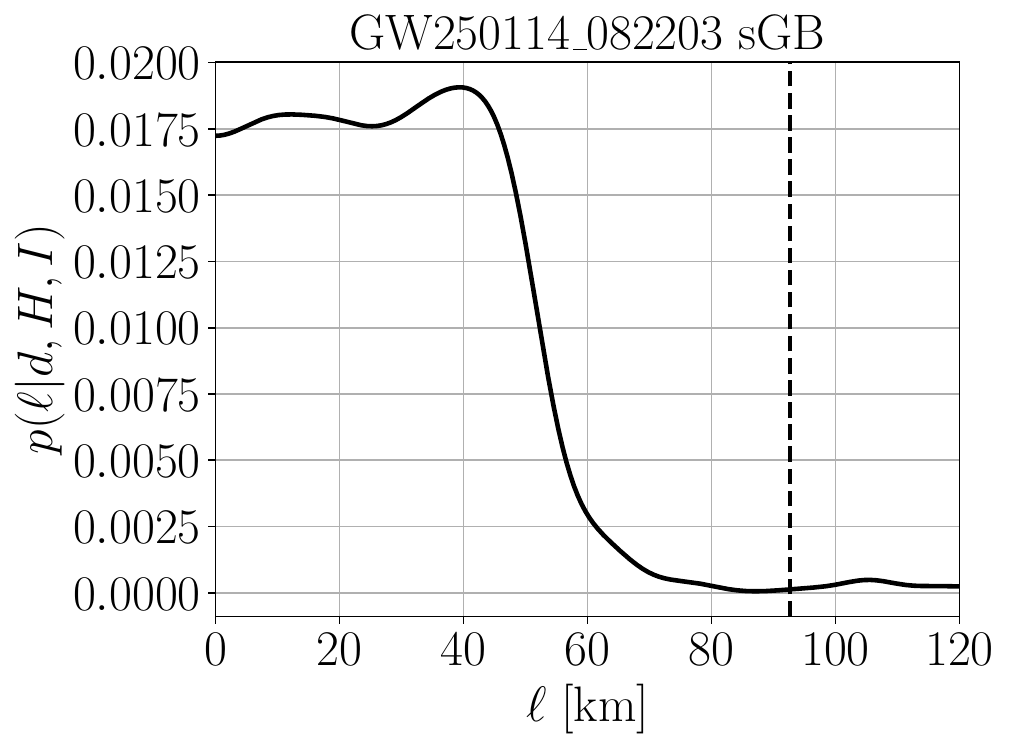}}
\quad
\subfloat{\includegraphics[width=0.25\textwidth]{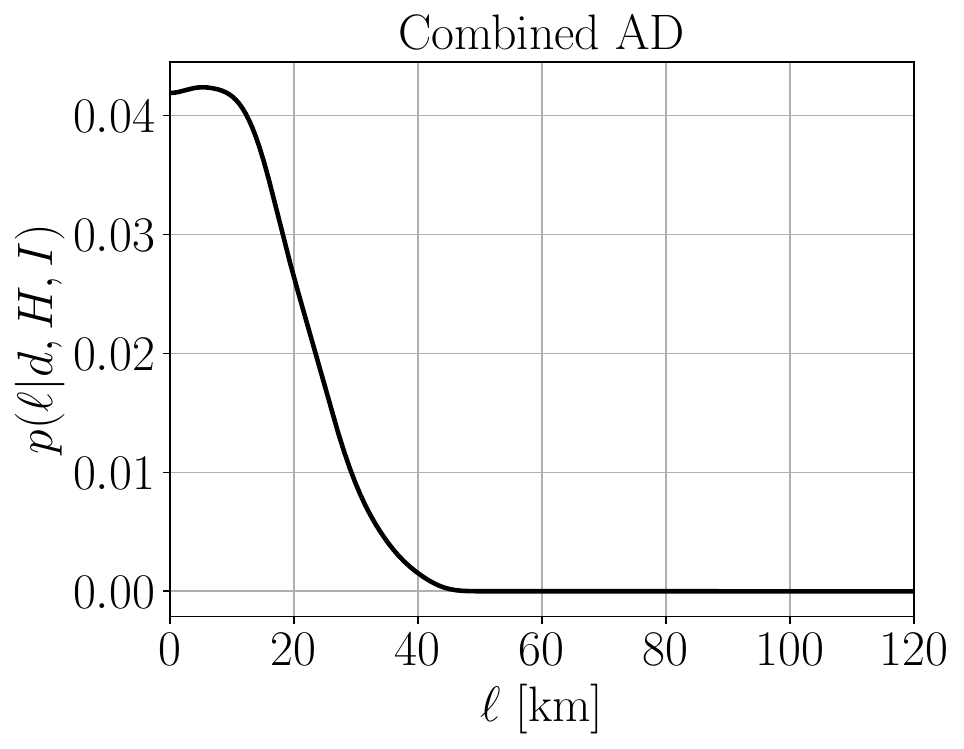}}
\subfloat{\includegraphics[width=0.25\textwidth]{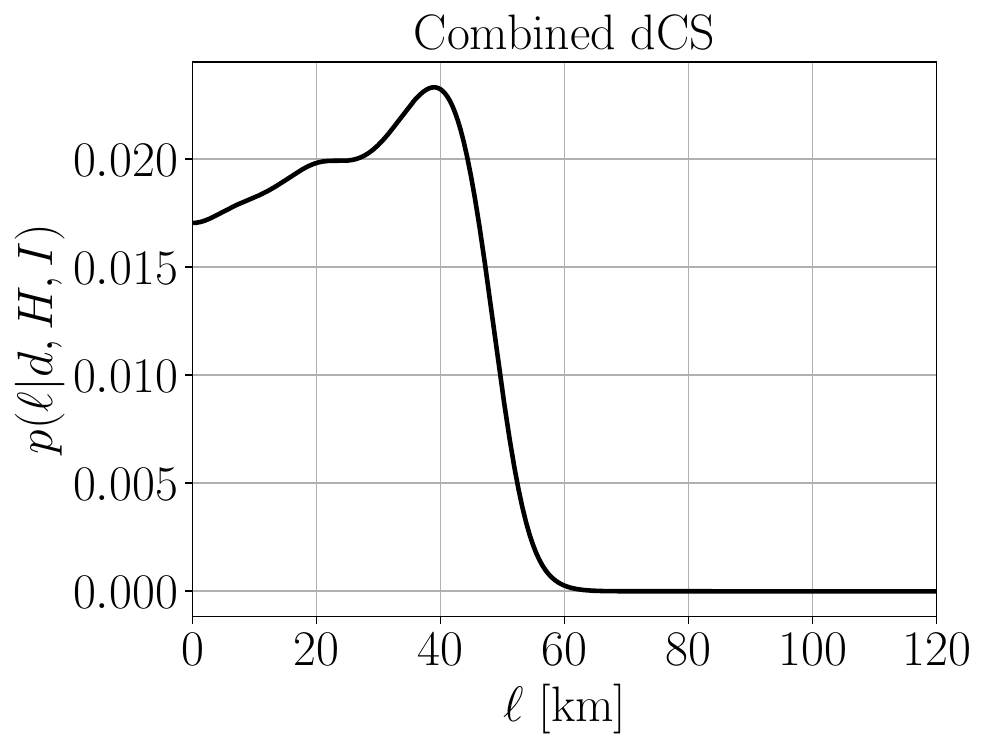}}
\subfloat{\includegraphics[width=0.25\textwidth]{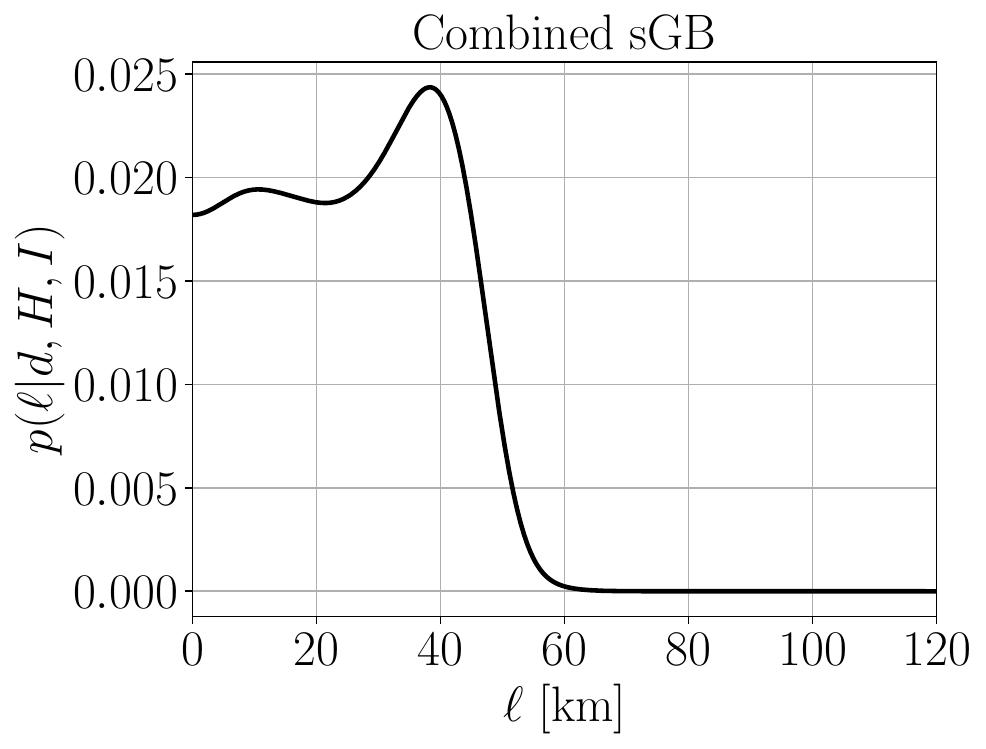}}
\caption{Marginalized posterior of $\ell$ of the scalar field(s) in axi-dilaton (AD, left column), dynamical Chern-Simons (dCS, middle column), and scalar Gauss-Bonnet (sGB, right column) gravity obtained by Bayesian inference of the ringdown signal of GW150914 (top column), GW190521\_074359 (second column), GW200129\_065458 (third column), GW231226\_101520 (fourth column), GW250114\_082203 (fifth column), and combined across all events (bottom column). 
In panels concerning individual ringdown signals, the dashed vertical lines indicate the maximum allowed $\ell := M_{\rm map}$ within the small-coupling approximation for the corresponding signal, where $M_{\rm map}$ is the maximal-posterior remnant mass of the individual ringdown signals in the source frame inferred assuming $\ell=0$. 
}
\label{fig:PDF_event_by_event}
\end{figure*}

Figure~\ref{fig:PDF_event_by_event} shows the marginalized posterior distribution on $\ell$ obtained by Bayesian inference of individual gravitational-wave events. The model is as given in Eq. (4) of the main text, and the parameters varied in our Bayesian analysis are $\{M_z,a,|A^{\pm,\rm (P)}_{022}|,\phi^{\pm,\rm (P)}_{022},\ell, d_L \}$, where $t_s = 10 M$ for the production runs (and then varied for robustness checks below). 
The priors for these parameters were described in the previous section. 
Observe that that all marginalized posterior distributions in Fig.~\ref{fig:PDF_event_by_event} are within the small coupling approximation, i.e.~the 90\% posterior weight is inside the small-coupling threshold indicated (vertical dashed lines). This consistency with the small-coupoing approximation will be further validated below.
Observe also that the 90\% constraints on $\ell$ are all better than $60$--$80$km, so below the horizon scale for all remnant black holes, since $r_+ \sim 2 M \approx \{183,419,187,214, 188\}$km for the GW150914, GW190521\_074359, GW200129\_065458, GW231226\_101520, and GW250114\_082203 events respectively. 
Finally, observe also that the constraint on $\ell$ in axi-dilaton gravity is the strongest among the three extensions. 
This is reasonable because the shift of the quasinormal-mode frequencies in axi-dilaton gravity is the sum of that in dynamical Chern-Simons and scalar Gauss-Bonnet gravity theories. 
In the former, axial quasinormal-mode spectra are more significantly modified, while in the latter, stronger modifications appear in the polar spectra. 
Consequently, axi-dilaton coupling modifies both parity sectors manifestly, making it an easier theory to rule out and leading to the most stringent constraints on $\ell$.

\section{Quasinormal-mode spectra of black holes coupled to additional fields}

\subsection{Lagrangian density}

The massless (pseudo)scalar fields searched in this work are governed by the following Lagrangian density, 
\begin{equation}\label{eq:Lagrangian}
\begin{split}
16 \pi \mathscr{L} = & R + \ell_1{}^2 \varphi_1 \mathscr{G} + \ell_2{}^2 \varphi_2  \mathscr{P} \\
& - \frac{1}{2} \nabla_{\mu} \varphi_1 \nabla^{\mu} \varphi_1 - \frac{1}{2} \nabla_{\mu} \varphi_2 \nabla^{\mu} \varphi_2.  
\end{split}
\end{equation}
A more general form of the coupling functions (the coefficient of $\mathscr{G}$ and $\mathscr{P}$) can, in principle, be considered.
In this work, we focus on the chosen form because it serves as the leading-order approximation of more complex coupling functions in the regime where $\alpha_{1,2}$ and $|\varphi_{1,2}|$ are small.
This is a justified assumption given that general relativity has passed all tests. 
Note that changing the sign of $\ell_{1, 2}$ does not change the Lagrangian density. 
Thus, in this work, we shall focus only on the case of $\ell_{1, 2}>0$. 

Through coupling to gravity, the fields can alter gravitational waves emitted by binary black-hole coalescence.
To model these changes, we need to solve the field equations governing the space-time $g_{\mu \nu}$ and the fields $\varphi_{1, 2}$, which can be derived by varying the Lagrangian density with respect to $g_{\mu \nu}$ and $\varphi_{1, 2}$. 
For the ease of keeping track of dimensions, we define the rescaled fields $\vartheta_{1,2} = \ell_{1, 2}{}^{-2} \varphi_{1, 2}$ and the dimensionless coupling parameter $\zeta_{1,2} = (\ell_{1,2} / M)^4$, where $M$ is the black-hole mass. 
In terms of the rescaled fields and the dimensionless coupling parameters, the field equations are
\begin{align}
& R_{\mu}{}^{\nu} + \sum_{q=1,2} \zeta_q \left[ \left( \mathscr{A}_q \right)_{\mu}{}^{\nu} -  \left( T_q \right)_{\mu}{}^{\nu} \right] = 0, \label{eq:field_eqs}\\
& \Box \vartheta_1+\mathscr{G} = 0, ~~~~ \Box \vartheta_2+\mathscr{P} = 0. 
\end{align}
Here, $\left( \mathscr{A}_{q=1,2} \right)_{\mu}{}^{\nu}$ are rank-2 tensors that are defined through the curvature tensor and the derivative of $\vartheta_{1, 2}$,
\begin{equation}
\begin{split}
\left( \mathscr{A}_1 \right)_{\mu}{}^{\nu}  = & ~ \delta^{\nu \sigma \alpha \beta}_{\mu \lambda \gamma \delta} R^{\gamma \delta}{}_{\alpha \beta}\nabla^{\lambda} \nabla_{\sigma} \vartheta_1  \\
& - \frac{1}{2} \delta_{\mu}{}^{\nu} \delta^{\eta \sigma \alpha \beta}_{\eta \lambda \gamma \delta} R^{\gamma \delta}{}_{\alpha \beta}\nabla^{\lambda} \nabla_{\sigma} \vartheta_1, \\
\left( \mathscr{A}_2 \right)_{\mu}{}^{\nu} = & - 4 g_{\mu \beta} \nabla_{\rho} \nabla_{\sigma}\left[ \tilde{R}^{\rho (\beta \nu) \sigma}\vartheta_2 \right], \\
\end{split}
\end{equation}
where $\delta^{\eta \sigma \alpha \beta}_{\eta \lambda \gamma \delta}$ is the generalized Kronecker delta symbol, whose definition is available as, for example, Eq. (13) of \cite{Chung:2024vaf}. 
$\left( T_q \right)_{\mu}{}^{\nu}$ is the energy-momentum tensor of the rescaled fields, 
\begin{equation}
\begin{split}
\left( T_q \right)_{\mu}{}^{\nu} & = \frac{1}{2} \nabla_{\mu} \vartheta_q \nabla^{\nu} \vartheta_q. 
\end{split}
\end{equation}
$\Box = (-g)^{-\frac{1}{2}}\partial_{\gamma} \left((-g)^{\frac{1}{2}}  g^{\gamma \lambda} \partial_{\lambda}\right)$ is the d'Alembertian operator. 
Note that $\varphi_2$ (or, equivalently, $\vartheta_2$) is a pseudo-scalar. 
Both $\vartheta_2$ and $\tilde{R}^{\alpha \beta \gamma \delta}$ pick up a minus sign upon parity transformation. 
Thus, the deformation from the black-hole space-time due to $\varphi_2$ still respects parity symmetry.

\subsection{Shift of the quasinormal-mode spectra}

By perturbatively solving the field equations in a black-hole background, we can compute the shift of the quasinormal-mode spectra. 
As the quasinormal-mode spectra in dynamical Chern-Simons and scalar Gauss-Bonnet gravity have been well studied \cite{Chung:2024vaf, Chung:2024ira, Chung:2025gyg}, we shall focus on axi-dilaton gravity here. 
We first formulate the field equations without explicitly involving the additional fields by writing the fields as 
\begin{equation}
\vartheta_{1} = - \Box^{-1} \mathscr{G}, ~~~ \vartheta_{2} = - \Box^{-1} \mathscr{P}, 
\end{equation}
where $\Box^{-1}$ is the inverse of the d'Alembertian operator. 
Substituting these expressions back into the tensor equations, we can rewrite $\left( \mathscr{A}_{q} \right)_{\mu}{}^{\nu}$ and $\left( T_q \right)_{\mu}{}^{\nu}$ as 
\begin{equation}
\begin{split}
\left( \mathscr{A}_{1} \right)_{\mu}{}^{\nu} = & - \delta^{\nu \sigma \alpha \beta}_{\mu \lambda \gamma \delta} R^{\gamma \delta}{}_{\alpha \beta}\nabla^{\lambda} \nabla_{\sigma} \left( \Box^{-1} \mathscr{G} \right) \\
& + \frac{1}{2} \delta_{\mu}{}^{\nu} \delta^{\eta \sigma \alpha \beta}_{\eta \lambda \gamma \delta} R^{\gamma \delta}{}_{\alpha \beta}\nabla^{\lambda} \nabla_{\sigma} \left( \Box^{-1} \mathscr{G} \right), \\
\left( \mathscr{A}_{2} \right)_{\mu}{}^{\nu} = & 4 g_{\mu \beta} \nabla_{\rho} \nabla_{\sigma}\left[ \tilde{R}^{\rho (\beta \nu) \sigma} \Box^{-1} \mathscr{P} \right], \\
\left( T_1 \right)_{\mu}{}^{\nu} = & \frac{1}{2} \nabla_{\mu} \left( \Box^{-1} \mathscr{G} \right) \nabla^{\nu} \left( \Box^{-1} \mathscr{G} \right), \\
\left( T_2 \right)_{\mu}{}^{\nu} = & \frac{1}{2} \nabla_{\mu} \left( \Box^{-1} \mathscr{P} \right) \nabla^{\nu} \left( \Box^{-1} \mathscr{P} \right).
\end{split}
\end{equation}

By solving the tensor field equations, we can construct the background space-time of a stationary black hole in axi-dilaton gravity. To first order in $\zeta$, we can decompose the metric tensor as 
\begin{equation}
g_{\mu \nu} = g^{\rm (0)}_{\mu \nu} + \zeta \; g^{\rm (1)}_{\mu \nu}, 
\end{equation}
where $g^{\rm (0)}_{\mu \nu}$ stands for the Kerr metric in general relativity and $g^{\rm (1)}_{\mu \nu}$ is the axi-dilaton deformation.
In \cite{Cano_Ruiperez_2019}, $g^{\rm (1)}_{\mu \nu}$ is solved for as a power series in $a$, $r^{-1}$ and $\chi$. 
For axi-dilaton gravity, $g^{\rm (1)}_{\mu \nu}$ is just the sum of that in dynamical Chern-Simons and scalar Gauss-Bonnet gravity simply because non-linear interactions between the scalars $\vartheta_{1}$ and $\vartheta_{2}$ lead to quadratic (and higher-order) terms in $\zeta$. 

Let us now consider quasinormal-mode metric perturbations in the Boyer-Lindquist coordinates $x^{\mu} = (t, r, \chi = \cos \iota, \phi)$. 
In these coordinates, the perturbed metric can be written as 
\begin{equation}\label{eq:metric_pert}
g_{\mu \nu} = g^{\rm GR}_{\mu \nu} + \zeta \; g^{\rm (1)}_{\mu \nu} + \varepsilon \; e^{i m \phi - i \omega t} \hat{h}_{\mu \nu} (r, \chi), 
\end{equation}
where $\varepsilon \ll 1$ is a bookkeeping parameter for the perturbations, $m$ and $\omega$ are the magnetic mode number and the complex frequency of the perturbations.
Let us denote the perturbations of $N_{\zeta}$-th order in $\zeta$ and $N_{\varepsilon}$-th order in $\epsilon$ to a rank-2 tensor with a superscript $(N_{\zeta}, N_{\varepsilon})$. 
In this notation, we see that 
\begin{equation}\label{eq:X}
\left[R_{\mu}{}^{\nu} \right]^{(1, 1)} + \sum_{q=1,2} \zeta_q \left[ \left( \mathscr{A}_q \right)_{\mu}{}^{\nu} -  \left( T_q \right)_{\mu}{}^{\nu} \right]^{(0,1)} = 0,
\end{equation}
which immediately implies that metric perturbations at first-order in $\zeta$ is the sum of two parts, each of which is sourced by the linearization of the dynamical Chern-Simons terms $\left[ \left( \mathscr{A}_{2} \right)_{\mu}{}^{\nu} \right]^{(0,1)}$ and $\left[ \left( T_{2} \right)_{\mu}{}^{\nu} \right]^{(0,1)}$, and the scalar Gauss-Bonnet terms $\left[ \left( \mathscr{A}_{1} \right)_{\mu}{}^{\nu} \right]^{(0,1)}$ and $\left[ \left( T_{1} \right)_{\mu}{}^{\nu} \right]^{(0,1)}$  on the Kerr background. 
From this equation, it already follows that the leading-order-in-$\zeta$ frequency shift of the quasinormal-mode frequencies in axi-dilaton gravity is the sum of that in the two gravity theories. 

To be more specific, let us solve Eq.~\eqref{eq:X} using METRICS (Metric pErTuRbations wIth speCtral methodS). 
To simplify the problem, we enforce the Regge-Wheeler gauge, which many quadratic-gravity theories have sufficient gauge degrees of freedom to enforce \cite{Wagle:2019mdq}. 
In this gauge, the metric perturbations are completely specified by 6 unknowns, which we write $h_{j=1,2, ..., 6}$. 
Since $\hat{h}_{\mu \nu}$ is purely ingoing at the horizon and purely outgoing at spatial infinity, $h_j$ is diverging at these two boundaries. 
Before we perform spectral expansions on $h_j$, we need to construct an asymptotic factor $A_j (r)$ to accommodate this divergent behaviour. 
An explicit form of $A_j (r)$ can be specified via the surface gravity and angular velocity of the event horizon of the black hole. 
Using the $A_j (r)$ asymptotic factor, the $h_j$ functions admit a spectral expansion that can be written schematically as 
\begin{equation}\label{eq:spectral_expansion}
h_j = A_j (r) \sum_{p=1}^{N} \sum_{\ell=1}^{N} v_{j, p, \ell} ~ \varphi_{p, \ell} (r, \chi)\,,
\end{equation} 
where no sum on $j$ is implied, $\varphi_{p, \ell} (r, \chi)$ is a complete and orthogonal spectral basis of $r$ and $\chi$, and the $v_{j, p, \ell}$ are constants.

Substituting Eq.~(\ref{eq:spectral_expansion}) into Eq.~(\ref{eq:metric_pert}), and then substituting the result into Eq.~(\ref{eq:field_eqs}), we expand the tensor field equations to first order in $\varepsilon$ and obtain a system of homogeneous linear algebraic equations for the coefficients $v_{j,p,\ell}$. 
Let us define the vector $\mathbf{v}$ to collect all components of $v_{j,p,\ell}$, namely  $\mathbf{v} =\{v_{j,p,\ell}\}$.
With this notation, the algebraic system can be written as
\begin{equation}\label{eq:linalg_eqn}
\mathbb{D} (\omega) \cdot \mathbf{v} = \left[ \mathbb{D}^{(0)} (\omega) + \zeta \, \mathbb{D}^{(1)} (\omega) \right] \cdot \mathbf{v} = \mathbf{0} \,,
\end{equation}
where $\mathbb{D}(\omega)$ is a rectangular matrix of dimension $10 N^2 \times 6 N^2$, whose entries depend on the frequency $\omega$ and background parameters $M_z$, $a$, and $M_z$.
Since Eq.~\eqref{eq:linalg_eqn} is valid only to first order in $\zeta$, we seek solutions that are also accurate to this order. 
To proceed, we specify one component of $\mathbf{v}^{(0)}$ and $\mathbf{v}^{(1)}$, as in \cite{Chung:2023wkd,Chung:2024vaf}. 
The remaining undetermined components of $\mathbf{v}^{(0)}$ ($\mathbf{v}^{(1)}$), together with $\omega^{(0)}$ ($\omega^{(1)}$), are collected into vectors $\mathbf{x}^{(0)}$ ($\mathbf{x}^{(1)}$) respectively.
Applying the spectral perturbation theory developed in METRICS, we obtain
\begin{equation}\label{eq:Eigenval_pert}
\mathbf{x}^{(1)} = - \mathbb{J}^{-1} \cdot \left( \left. \mathbb{D}^{(1)} \right|_{\omega^{(0)}} \cdot \mathbf{v}^{(0)} \right) \,,
\end{equation}
where $\mathbb{J} = \partial_{\mathbf{x}^{(0)}} \mathbb{D}^{(0)}$ is the Jacobian matrix, and $\mathbb{J}^{-1}$ denotes its generalized inverse.
The first-order-in-$\zeta$ terms in Eq.~\eqref{eq:field_eqs} consist of a group terms that follow from dynamical Chern-Simons gravity and another group from scalar Gauss-Bonnet gravity. 
Naturally, we can then write 
\begin{eqnarray}
\mathbb{D}^{(1)} (\omega) = \sum_{q=1,2} \mathbb{D}^{(1)}_{q} (\omega), 
\end{eqnarray}
where $\mathbb{D}^{(1)}_{q} (\omega)$ is the coefficient matrix obtained from performing an spectral expansion on $\left[ \left( \mathscr{A}_{q} \right)_{\mu}{}^{\nu} \right]^{(0,1)}$ and $\left[ \left( T_{q} \right)_{\mu}{}^{\nu} \right]^{(0,1)}$. 
Then, it follows from Eq.~\eqref{eq:Eigenval_pert} that 
\begin{equation}
\omega^{(1)} = \sum_{q=1,2} \omega^{(1)}_{q}, 
\end{equation}
which means that the lead-order-in-$\zeta$ shift of quasinormal-mode spectra due to the additional field in axi-dilaton gravity is the sum of that due to the individual fields. 
From this equation Eq.~(2) in the main text immediately follows. 


\section{Further robustness checks of the results}

In this section, we perform multiple checks of robustness and of the correctness of our results. 
As observed from Fig.~(\ref{fig:PDF_event_by_event}), the marginalized $\ell$ posterior of different additional fields inferred from different signals is similar. 
Thus, except for the assessment of parameter degeneracies, we will focus on dynamical Chern-Simons coupling using the ringdown signal of GW150914 as an example; we expect qualitatively similar conclusions to hold when analyzing different ringdown signals and couplings.

\subsection{Consistency with the small-coupling approximation}

\begin{figure*}[htp!]
\centering  
\subfloat{\includegraphics[width=0.47\linewidth]{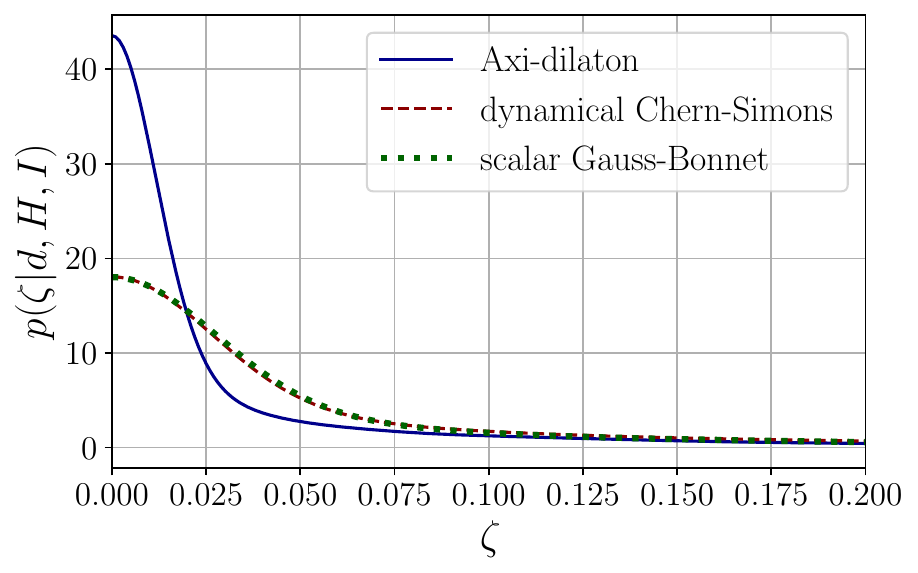}}
\subfloat{\includegraphics[width=0.47\linewidth]{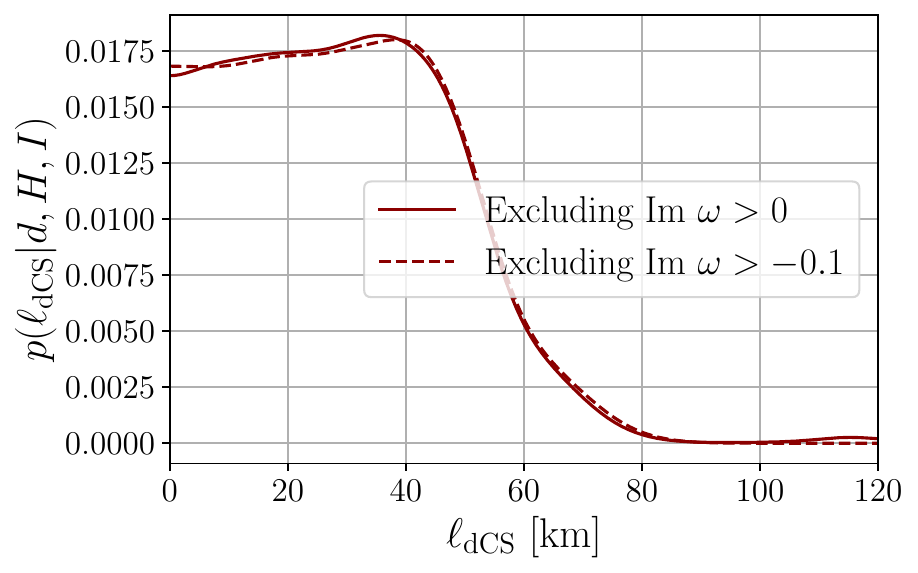}}
\caption{(Left panel) Marginalized posterior distribution of $\zeta$, obtained by converting the samples of $d_L$, $M_z$ and $\ell$ inferred from the GW150914 ringdown signal via Eq.~(3) in the main text. 
We only show the posterior up to $\zeta = 0.2$, as the posterior has only negligible support for $\zeta > 0.2$. 
Observe that most of the $\zeta$ samples are within $0.2$, indicating the close consistency between the results of our analyses with the small-coupling approximation. 
(Right panel) Marginalized posterior distribution of $\ell$, inferred from the GW150914 ringdown signal, using priors that exclude $\text{Im} (\omega) > 0$ (solid blue) and $\text{Im}(\omega) > -0.1$ (dashed blue). 
Observe that the two posteriors are nearly identical, indicating that the results are not sensitive to the exact $\text{Im}(\omega)$ cut-off. 
We conclude that the posterior of $\ell$ is not driven by the exclusion of $\zeta > 1$ or $\text{Im}(\omega) > 0$ from the prior.
}
\label{fig:zeta}
\end{figure*}

The left panel of Fig.~(\ref{fig:zeta}) shows the marginalized posterior of $\zeta$ as a function of $\ell$ for axi-dilaton (solid), dynamical Chern-Simons (dashed), and scalar Gauss-Bonnet (dotted) couplings, derived from samples of $M_z$, $d_L$, and $\ell$ using Eq.~(3) in the main text, based on the GW150914 ringdown signal. 
In the interest of space, we only show the posteriors of $\zeta < 0.2$, as the posteriors beyond this range have negligible support. 
Observe that most of the samples lie within the $\zeta \lesssim 0.2$ region. Note also that $\zeta \lesssim \mathcal{O}(10^{-1})$ denotes the regime of coupling parameter space that is consistent with the small-coupling approximation. This justifies the use of the METRICS quasinormal-mode frequencies, which are only valid up to first order in $\zeta$. 
Furthermore, the posterior of $\zeta$ shows that the posterior of $\ell$ is not an artifact of excluding values of $\ell$ that lead to $\zeta > 1$. 

We also verified that the posterior is not affected by the exclusion of values of $\ell$ in the prior that lead to $\text{Im}  (\omega) > 0$ (see discussion in the last paragraph of the parameter estimation section). 
The right panel of Fig.~\ref{fig:zeta} shows the marginalized posterior of $\ell$ of dynamical Chern-Simons coupling inferred from the GW150914 ringdown signal using two priors: one excluding $\text{Im} (\omega) > 0$ (solid blue) and another excluding $\text{Im} (\omega) > -0.1$ (dashed blue). 
Observe that the two posteriors overlap almost completely, indicating that adjusting the exclusion region of $\text{Im} ( \omega)$ within a reasonable range does not significantly affect our constraints on $\ell$. 
The 90-\% confidence levels on $\ell$ obtained by excluding $\text{Im} (\omega) > 0 $ and $\text{Im} (\omega) > -0.1$ are 53.6 km and 53.7 km, respectively.
In other words, our posteriors are not sensitive to the precise cut-off chosen for $\text{Im}(\omega)$. 

\subsection{Degeneracy amongst $\ell, M$ and $a$}

\begin{figure*}[htp!]
\centering  
\subfloat{\includegraphics[width=6cm]{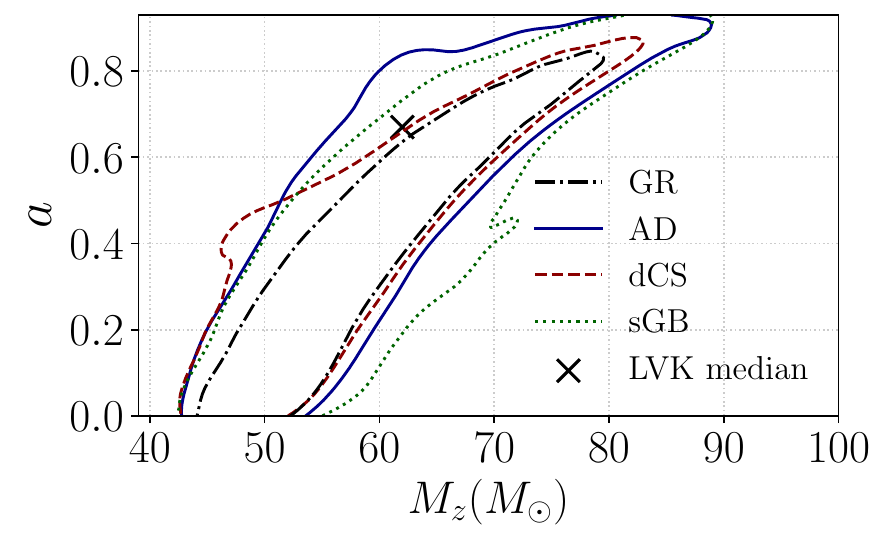}}
\subfloat{\includegraphics[width=6cm]{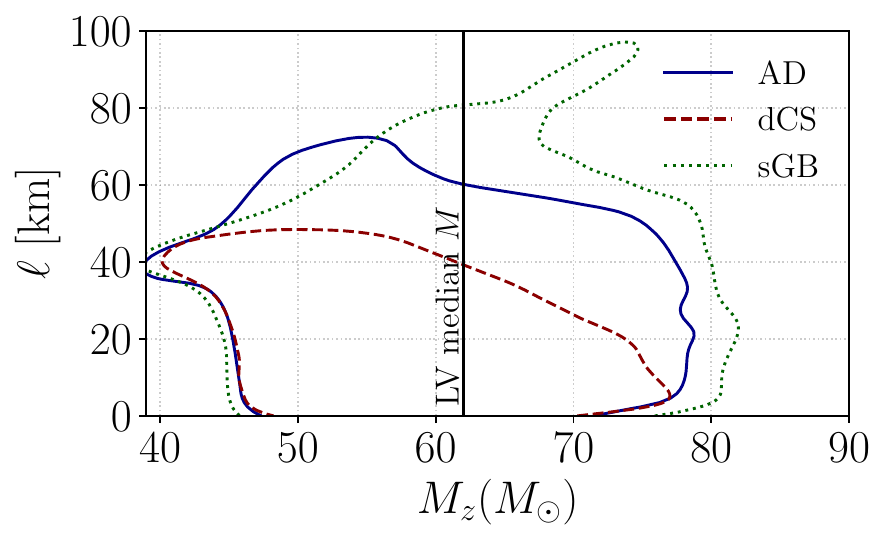}}
\subfloat{\includegraphics[width=6cm]{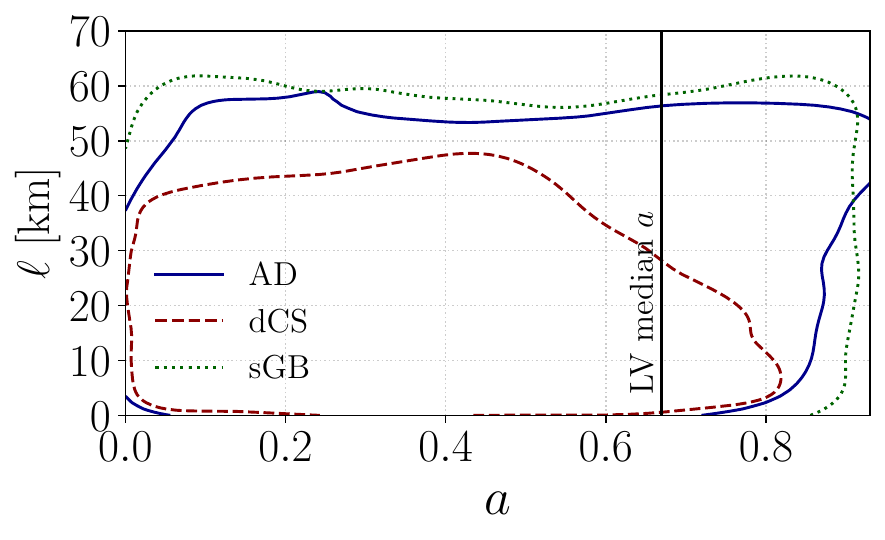}}
\caption{The 90\% credible region of the marginalized two-dimensional posteriors in the $M_z$–$a$ (left), $M_z$–$\ell$ (middle), and $a$–$\ell$ (right) planes, inferred from the ringdown phase of GW150914 assuming axi-dilaton (solid blue), dynamical Chern–Simons (dashed red), scalar Gauss–Bonnet (dotted green) couplings, and general relativity ($\ell_1 = \ell_2 = 0$, dashed-dotted black, shown only in the left panel). 
Observe that the general-relativity contour lies very close to the median remnant mass and spin reported by the LIGO–Virgo–KAGRA collaboration (marked by $\times$, labeled “LVK median”), validating our general-relativity ringdown analysis. 
Observe also that the inclusion of $\ell$ does not induce significant distortion to the contours, implying that there should be no strong correlations between $\ell$ and either $M_z$ or $a$.  
The deduction is further consolidated by the observation that the contours in the $M_z-\ell$ and $a-\ell$ planes exhibit no significant inclination. 
}
\label{fig:Denegrancy}
\end{figure*}

Fig.~(\ref{fig:Denegrancy}) shows the 90\% credible region of the two-dimensional marginalized posteriors in the $M_z$–$a$ (left), $M_z$–$\ell$ (middle), and $a$–$\ell$ (right) planes, obtained from the analysis of the ringdown phase of GW150914 assuming: general relativity ($\ell_1 = \ell_2 = 0$, solid black, shown only in the left panel), axi-dilaton (solid blue), dynamical Chern–Simons (dashed red), and scalar Gauss–Bonnet (dash-dotted green) couplings.
In the left panel, observe that the general relativity posterior is consistent with previous studies \cite{pyRing_03, Cheung:2020dxo} and it lies close to the LIGO-Virgo-KAGRA median values \cite{LIGOScientific:2016vlm}, marked by $\times$), confirming the validity of our general-relativity inference. 
The posteriors obtained using quadratic gravity models are not significantly distorted, suggesting no strong degeneracy between $\ell$, $M_z$, and $a$.
This conclusion is supported by the middle and right panels. 
The vertical solid black lines indicate the LIGO–Virgo–KAGRA median remnant mass (middle panel) and spin (right panel). 
All credible regions show no significant degeneracies in the $M_z$–$\ell$ or $a$–$\ell$ planes, confirming that $\ell$ is not strongly correlated with the remnant properties.
We thus conclude that our constraints on $\ell$ are robust and do not compromise the accuracy of the inferred remnant mass and spin.

\subsection{Convergence of Bayesian inference}

\begin{figure}[htp!]
\centering  
\subfloat{\includegraphics[width=\columnwidth]{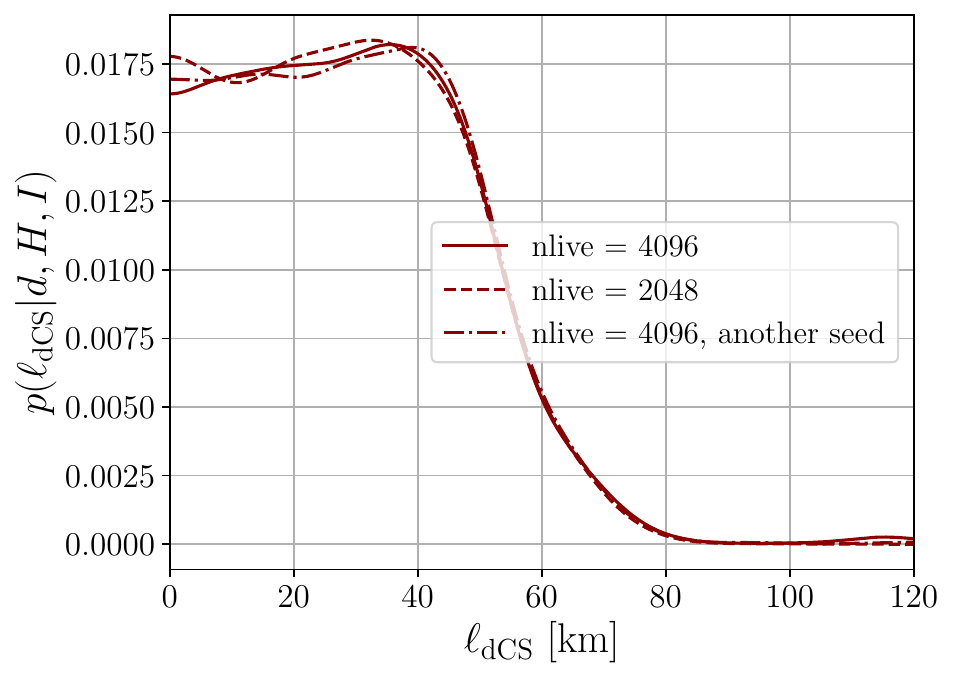}}
\caption{The marginalized posterior distribution of the coupling parameter $\ell$ for dynamical Chern–Simons coupling sampled from the ringdown signal of GW150914. 
The results are obtained using nested sampling with 4096 live points for two different random seeds (solid and dotted lines) and with 2048 live points (dashed line). 
All three posteriors are in good agreement with each other, demonstrating the robustness and convergence of the nested sampling procedure.
}
\label{fig:Nlive_convergence}
\end{figure}

Fig.~(\ref{fig:Nlive_convergence}) shows the posterior of $\ell$ for dynamical Chern-Simons coupling inferred from the GW150914 ringdown, using nested sampling with 4096 live points for two random seeds (solid and dashed) and 2048 live points (dotted). 
The 90-\% credible upper limits on $\ell$ of these three posteriors are 53.6 km, 53.5 km, and 52.9 km, respectively. The close agreement among all posteriors indicates the convergence of our Bayesian inference.

\subsection{The approximation of $A^{N, q}_{nlm} = (-1)^l (A^{P, q}_{nlm})^{\dagger}$}

\begin{figure}[htp!]
\centering  
\subfloat{\includegraphics[width=\columnwidth]{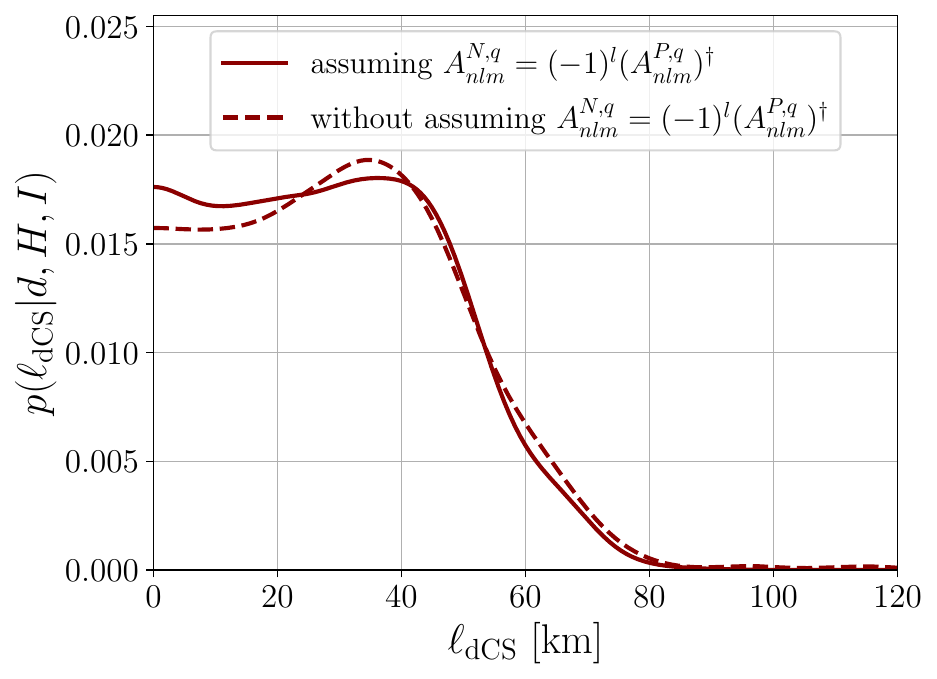}}
\caption{The marginalized posterior of $\ell$ for dynamical Chern-Simons coupling obtained by analyzing GW150914 ringdown signal, assuming (solid red) and without approximating (dashed red) $A^{N, q}_{nlm} = (-1)^l (A^{P, q}_{nlm})^{\dagger}$. 
Observe that the two posteriors overlay almost on top of each other, indicating that the constraint on $\ell$ is not significantly impacted by the approximation. 
}
\label{fig:Precession}
\end{figure}

The assumption $A^{N, q}_{nlm} = (-1)^l (A^{P, q}_{nlm})^{\dagger}$ is not required, but it reduces the number of free parameters and accelerates sampling. Fig.~(\ref{fig:Precession}) shows the posterior of $\ell$ for dynamical Chern-Simons coupling inferred from the GW150914 ringdown, with (solid) and without (dashed) imposing this relation. 
The 90-\% credible upper limits on $\ell$ obtained by assuming and not assuming $A^{N, q}_{nlm} = (-1)^l (A^{P, q}_{nlm})^{\dagger}$ are 53.7 km and 55.9 km, respectively. 
The two posteriors are nearly identical, indicating that the assumption has negligible impact on the constrain on $\ell$. 

\subsection{Robustness against the uncertainty in the METRICS frequencies}

\begin{figure*}[htp!]
\centering  
\subfloat{\includegraphics[width=0.47\linewidth]{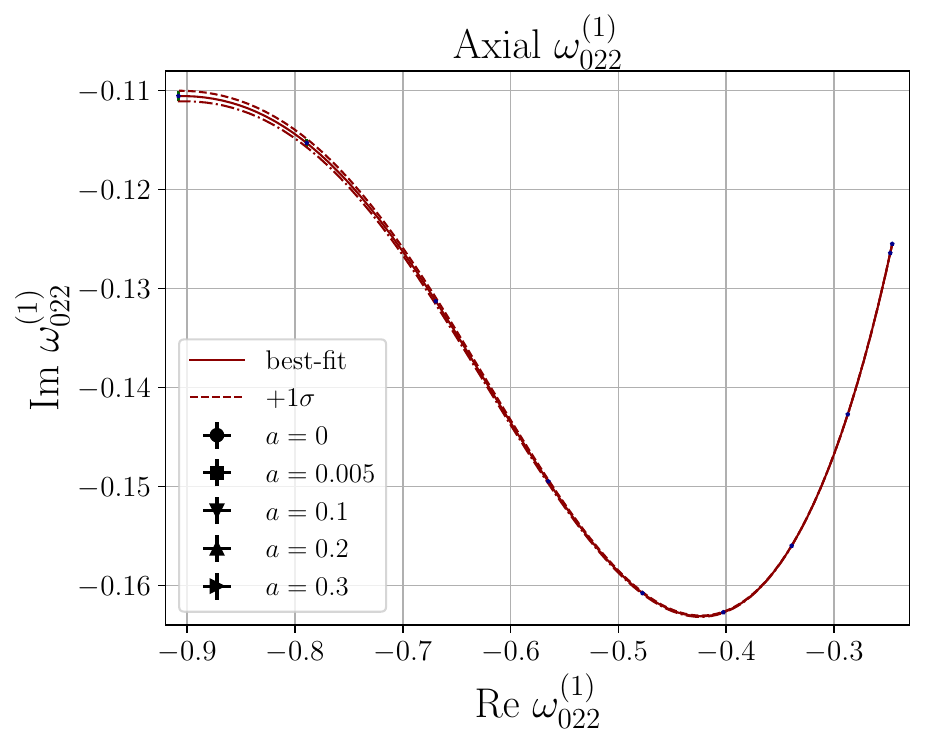}}
\subfloat{\includegraphics[width=0.47\linewidth]{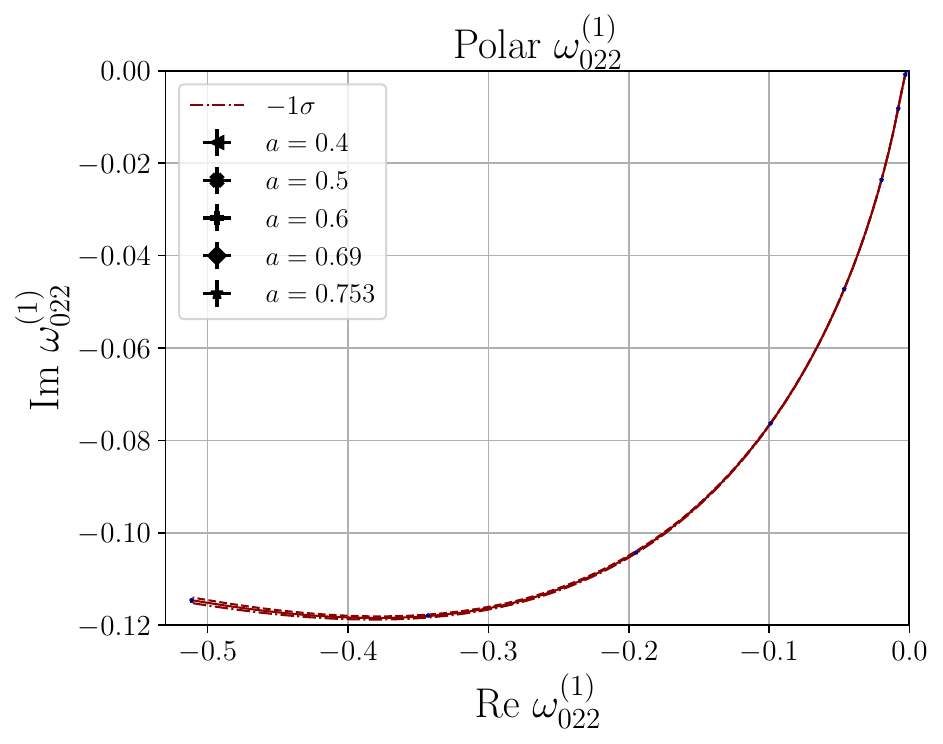}}
\caption{Trajectories in the complex plane traced by the METRICS 022-mode axial (left) and polar (right) frequencies of black holes subjected to dynamical Chern-Simons coupling. The different symbols correspond to the numerical frequencies calculated with METRICS with numerical error bars (small vertical lines in green, which are usually not visible because they are smaller than the symbol size) at different spin values. The solid, dashed and dot-dash curves correspond to the a polynomial fit to the frequencies, a fit to the frequencies shifted by $+1 \sigma$, and a fit to the frequencies shifted by $-1 \sigma$, respectively, where $\sigma$ is the $1\sigma$-numerical error of the frequencies at different spin values (c.f. Table. II of \cite{Chung:2025gyg}). 
Observe that the numerical error at small spin is much smaller than that at a large spin, and that the dashed and dot-dashed curves are very close to the solid curve. 
}
\label{fig:Spectra_w_uncertainty}
\end{figure*}

\begin{figure}[htp!]
\centering  
\subfloat{\includegraphics[width=\columnwidth]{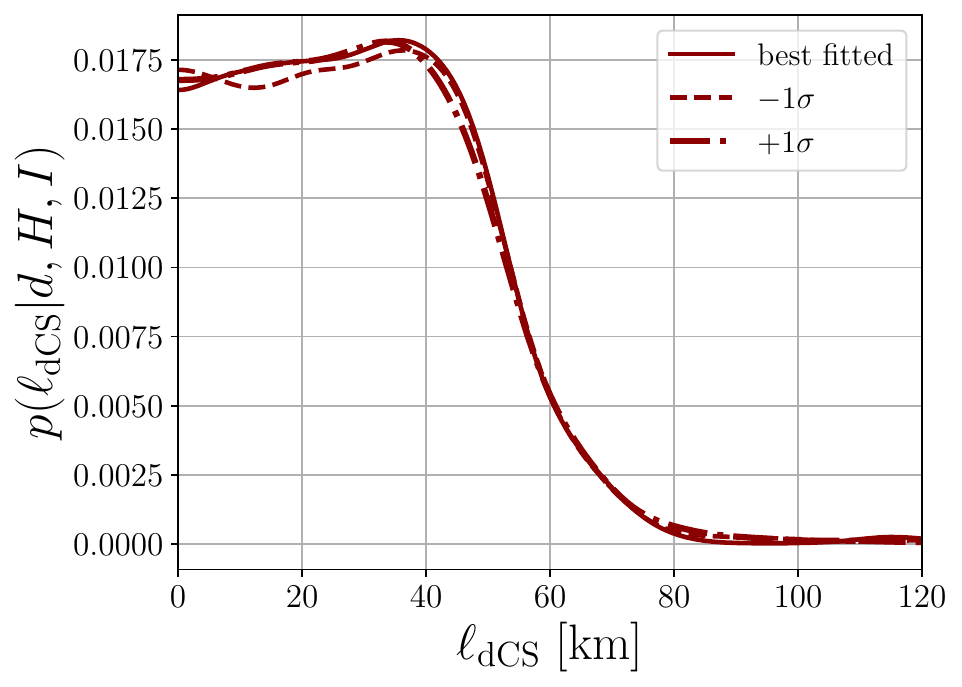}}
\caption{The marginalized posterior of $\ell$ for dynamical Chern–Simons coupling inferred from the GW150914 ringdown phase, using the METRICS 022-mode frequency at their best-fit values (solid), and shifted by $+1\sigma$ (dashed) and $-1\sigma$ (dotted). 
The close agreement among the posteriors demonstrates the robustness of our results against numerical uncertainties in the METRICS frequencies and their fitted coefficients.
}
\label{fig:Fitting_coeffs}
\end{figure}

The numerical errors in the METRICS computations lead to uncertainty in the fitted coefficients of the METRICS quasinormal-mode frequencies. 
The proper way to account for this uncertainty is to treat the coefficients as free parameters with a multivariate Gaussian prior, using the known variance-covariance matrix of the fitting coefficients. 
However, this approach significantly increases the number of inference parameters and the computational cost of nested sampling. 
To estimate the impact of numerical error of the METRICS frequencies, we construct a polynomial fit to the best-fit 022-mode frequencies shifted by $\pm 1 \sigma$, as shown in Fig.~(\ref{fig:Spectra_w_uncertainty}). 
Observe that at large spin, where the numerical error is the largest, the error is, at most, as large as the size of the scattered symbols.
After taking the numerical error in METRICS frequencies at all spins into account, the values of the frequencies computed using different fitting functions are not changed much at all.

Fig.~(\ref{fig:Fitting_coeffs}) presents the inferred posterior of $\ell$ for dynamical Chern-Simons coupling, using the fitting polynomial to the best-fit frequencies \cite{Chung:2025gyg} shifted by $+1\sigma$ (dashed) and $-1\sigma$ (dotted), which is to be compared with the posterior obtained using fitting expressions to the best-fit frequencies (solid). 
As shown in this figure, the posteriors are consistent with each other because, given the sensitivity of the existing detectors, the statistical error of the detection dominates the systematic errors induced by the numerical errors in the METRICS frequencies.
The 90-\% credible upper limits on $\ell$ obtained using the best-fit frequency, shifted by $+1\sigma$ and $-1\sigma$ is 53.6 km, 55.2 km and 54.7 km respectively. 
Thus, we conclude that, even if we infer the fitting coefficients as free parameters, the 90-\% credible upper limit will not be affected much at all. In other words, our results are robust against the numerical uncertainties in the METRICS frequencies. 

\subsection{Robustness against the systematic errors in the quasinormal-mode spectra model for spins $a > 0.75$}
\label{app:a<0.75}

\begin{figure*}[htp!]
\centering  
\subfloat{\includegraphics[width=0.47\linewidth]{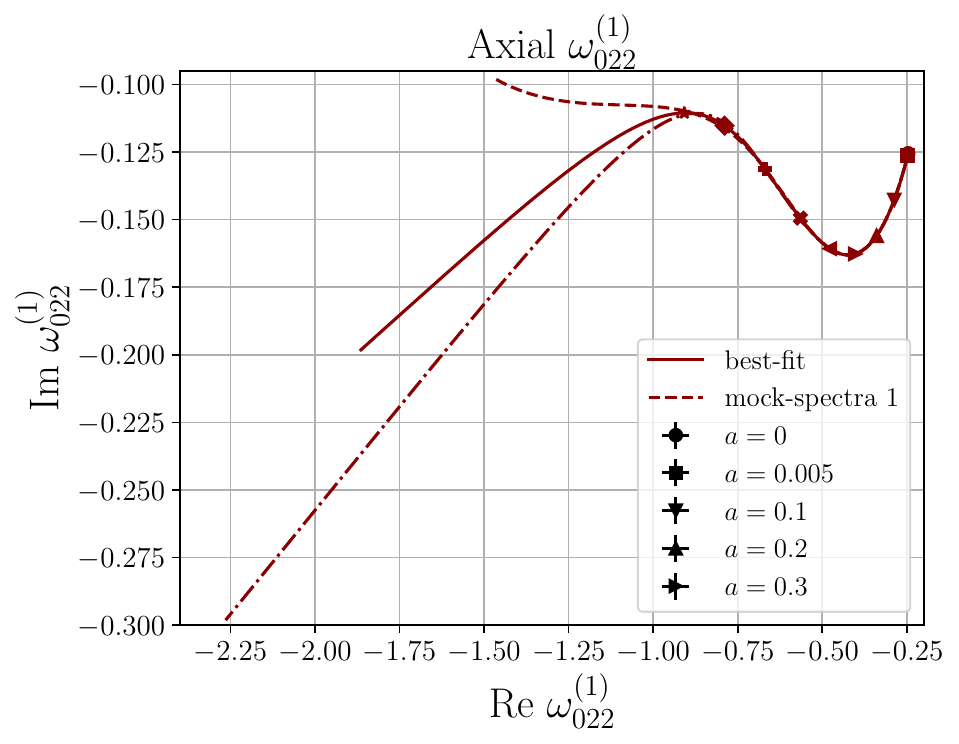}}
\subfloat{\includegraphics[width=0.47\linewidth]{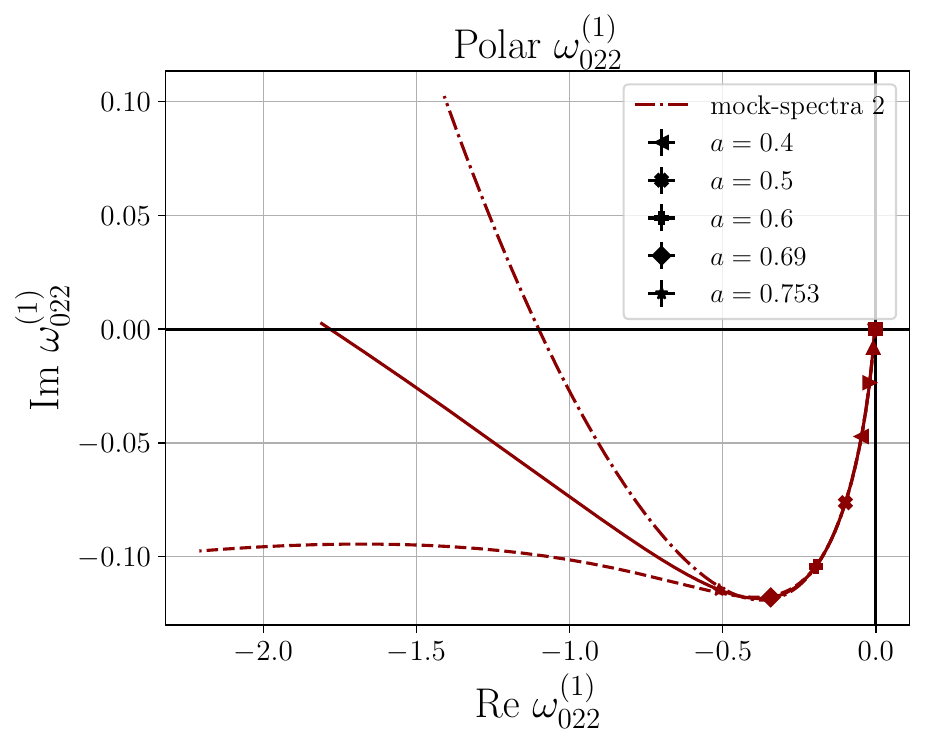}}
\caption{Quasinormal-mode frequencies of the axial (left) and polar (right) 022 modes, computed using the original fitting expressions to the METRICS data (solid lines) and two mock spectra (dashed and dash-dotted lines). 
The scatters mark frequencies computed from METRICS \cite{Chung:2025gyg}. 
For $a \leq 0.753$, all spectra show good agreement. Beyond this spin, the mock spectra diverge, spanning a broad region in the complex plane that represents the estimated uncertainty in the quasinormal-mode frequencies at high spin.
}
\label{fig:Mock_spectra}
\end{figure*}

As discussed, nested sampling occasionally explores regions with $a > 0.75$, where the quasinormal-mode spectra of black holes subjected to dynamical Chern–Simons (and hence axi-dilaton) coupling, computed using METRICS, have not been calculated accurately yet. 
To assess the impact of this limitation, we manufacture two mock-ups of possible 022-mode spectra as follows. 
First, we extrapolate the axial and polar 022-mode frequencies up to $a = 0.93$ using the existing fitting expressions. 
Then, we artificially displace both the real and imaginary parts of the frequencies at $a=0.93$ by $\pm 0.4$ and $\pm 0.1$, respectively. 
These displacements are chosen because we found that these are approximately the error of the corresponding parts of the Kerr 022-mode frequency at $a=0.93$, if one extrapolates the frequency from $a=0.753$ using the same number of frequencies and fitting polynomial of the same degree. 
We refer to these models as “mock spectra 1" and “mock spectra 2".
We fit each mock spectrum using a polynomial of one degree higher than the optimal degree used in the original fits. 
We check that the frequencies computed using the fitting expressions of these spectra are different from the mock data by only $\sim 10^{-3}$ for $a \leq 0.75$. 
Figure~\ref{fig:Mock_spectra} shows the axial (left) and polar (right) 022-mode frequencies in the complex plane, comparing the original fits (solid lines), the METRICS frequencies (scatters, taken from \cite{Chung:2025gyg}, and the two mock spectra (dashed and dash-dotted lines). 
All spectra agree well for $a \leq 0.753$, beyond which they diverge from each other, forming a broad spread in the complex plane that reflects the uncertainty in the quasinormal-mode spectrum at very large spins. 

\begin{figure}[tp!]
\centering  
\subfloat{\includegraphics[width=\columnwidth]{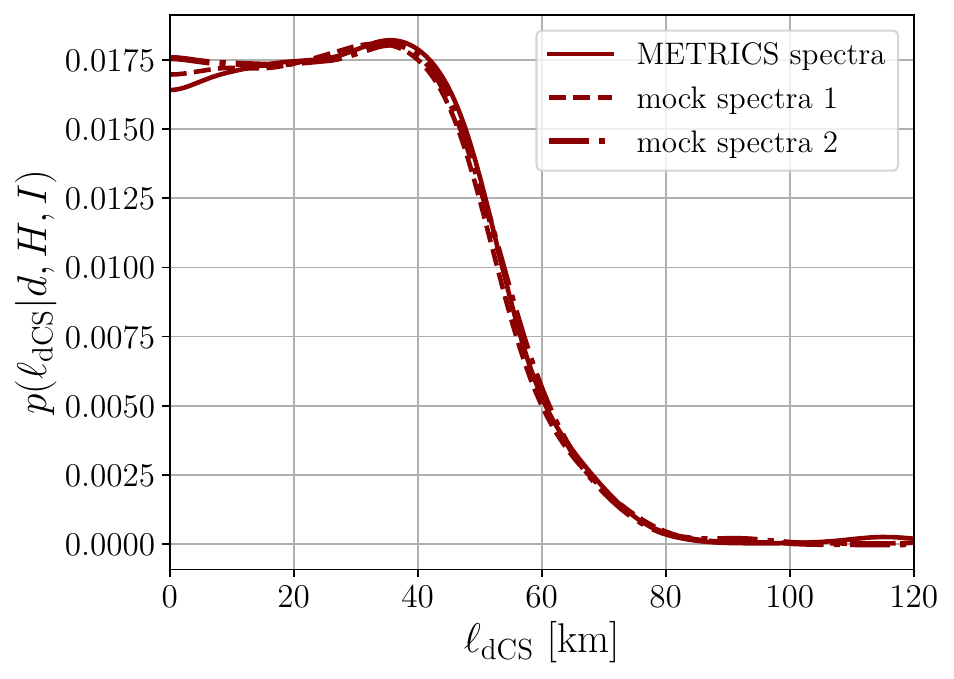}}
\caption{Marginalized posterior distribution of $\ell$ for dynamical Chern–Simons coupling inferred from the GW150914 ringdown signal, using the fitting expression for the METRICS frequencies (solid) and using the two mock spectra models (dashed and dash-dotted; see Fig.~\ref{fig:Mock_spectra} and the main text of this Appendix). 
The posteriors almost overlay on top of each other, indicating that our results are robust against the ignorance of the quasinormal-mode spectra for spins $a > 0.753$.
}
\label{fig:Ignorance}
\end{figure}

Fig.~(\ref{fig:Ignorance}) shows the marginalized posterior of $\ell$ for dynamical Chern–Simons coupling obtained using the original spectra (solid) and the two mock spectra models (dashed and dotted). 
Observe that the posteriors are largely consistent with each other.
The 90-\% credible upper limit on $\ell$ obtained using the METRICS spectra is $53.6$ km, while using the first and second mock spectra the 90\% credible upper limit is $54.2$ km and $53.4$ km, respectively. 
The latter two 90\% credible upper limits are not significantly affected regardless of the dependence of the quasinormal-mode spectra on $a > 0.753$, demonstrating the robustness of our results against the ignorance of the dependence of quasinormal-mode frequencies at $a > 0.75$. 
This robustness is expected, given that, as shown in Fig.~(\ref{fig:Denegrancy}), most of the posterior support lies inside $a < 0.75$.

\subsection{Robustness against uncertainty in the ringdown start time}

\begin{figure}[tp!]
\centering  
\subfloat{\includegraphics[width=\columnwidth]{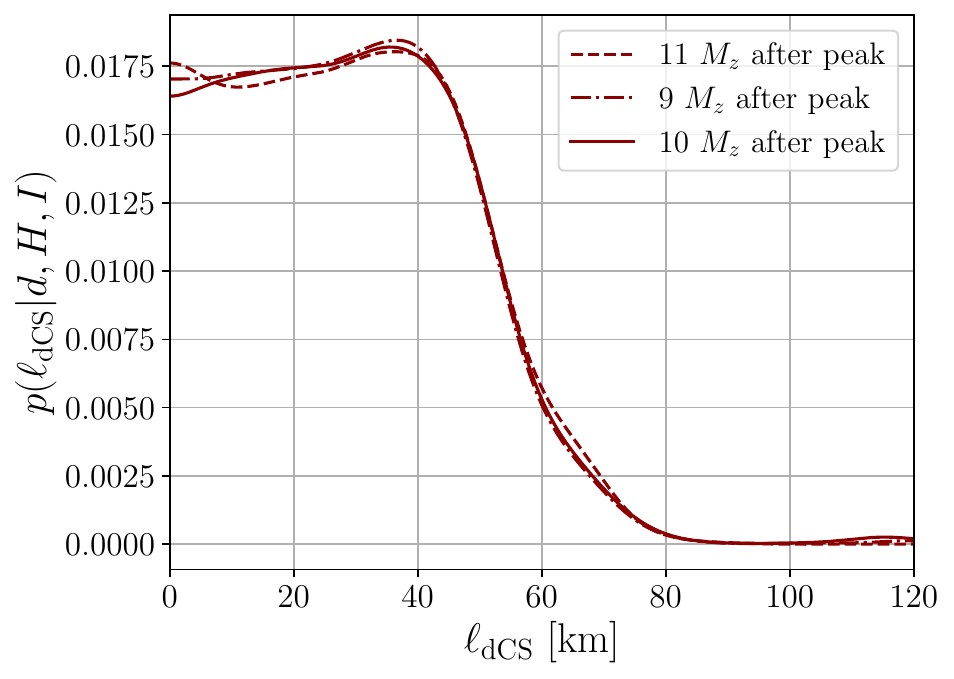}}
\caption{Marginalized posterior distribution of $\ell$ for dynamical Chern–Simons coupling inferred from the GW150914 ringdown signal, assuming the ringdown phase starts at $9M_z$ (dash-dotted), $10M_z$ (solid), and $11M_z$ (dashed) after the peak of $h_{+}^2 + h_{\times}^2$. Here, $M_z$ is the remnant mass estimated from a full waveform analysis in general relativity. The near-identical posteriors indicate that the results are robust against the choice of ringdown start time.
}
\label{fig:Time}
\end{figure}

We also checked that our results are robust against the choice of the ringdown start time. 
Fig.~(\ref{fig:Time}) shows the posterior of $\ell$ in dynamical Chern-Simons gravity infered from the GW150914 ringdown signal, assuming the ringdown signal starts at $9M_z$ (dashed-dotted), $10M_z$ (solid) and $11M_z$ (dotted) after the time when the amplitude $h_{+}^2 + h_{\times}^2$ is peaked. 
The 90-\% credible upper limits on $\ell$ obtained assuming the ringdown phase starts at $9 M_z$, $10 M_z$, and $11 M_z$ are 52.9 km, 53.6 km, and 53.7 km, respectively. All posteriors are consistent with each other, indicating that our results are robust against difference choices of the ringdown start time, provided that the non-linearity has been well subsided and a reasonable signal-to-noise ratio is able to be recovered by the choices. 

\subsection{Effects of the inclusion of an overtone}

\begin{figure}[htp!]
\centering  
\subfloat{\includegraphics[width=\columnwidth]{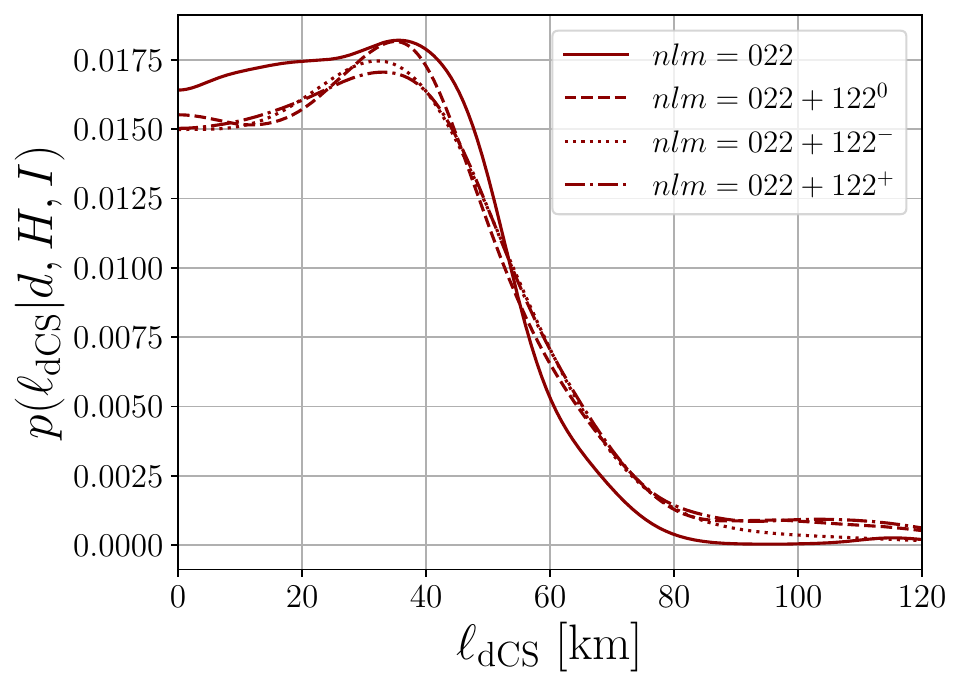}}
\caption{Marginalized posterior distribution of $\ell$ for dynamical Chern–Simons coupling inferred from the GW150914 ringdown phase, using the waveform model in Eq.~(4) in the main text with different mode combinations: the $nlm = 022$ mode alone (solid), the $022 + 122^{(0)}$ modes (dashed), the $022 + 122^{-}$ modes (dotted), and the $022 + 122^{+}$ modes (dash-dotted). 
The $122^{(0)}$ frequencies are estimated using the modified Teukolsky formalism, valid to first order in spin. 
The $122^{-}$ ($122^{+}$) modes are synthetic spectra constructed by decreasing (increasing) the real and imaginary parts of the $122^{(0)}$ frequencies at $a = 0.75$ by 100\%, linearly interpolated with the unmodified $a = 0$ value. 
The close agreement among all posteriors demonstrates that including an overtone does not substantially affect the inferred constraint on $\ell$, even with significant uncertainty in the overtone frequencies.
}
\label{fig:Overtone}
\end{figure}

To assess the impact of including an overtone, we infer the posterior of $\ell$ using both the $nlm = 022$ and $122$ modes. 
The overtone frequencies of black holes with $a \geq 0.75$ subjected to dynamical Chern–Simons coupling have yet to be accurately computed. 
The most up-to-date estimates come from the modified Teukolsky formalism, which is valid only to first order in spin. 
In our analysis here, we adopt the 122-mode frequencies computed using this formalism, denoted as the $nlm = 122^{0}$ spectrum, along with the METRICS 022-mode frequencies, to infer the posterior of $\ell$.
However, by comparing the METRICS frequencies (accurate to $a \sim 0.8$) with their first-order-in-spin counterparts, we find relative differences of up to 100\% \cite{Chung:2024vaf, Chung:2025gyg}. 
To account for this uncertainty, we construct two additional sets of synthetic 122-mode spectra. 
Specifically, we first evaluate the $122^{0}$ frequencies at $a = 0.75$, and then we artificially increase or decrease both their real and imaginary parts by 100\%, while keeping the values at $a = 0$ unchanged. 
We then perform a linear fit between the $a = 0$ values and the modified $a = 0.75$ points to create two new spectra: one with amplified values ($122^{+}$) and one with suppressed values ($122^{-}$).

Fig.~(\ref{fig:Overtone}) shows the marginalized posterior of $\ell$ for dynamical Chern–Simons coupling inferred using different mode combinations: the 022 mode alone (solid line), $022 + 122^{0}$ (dashed), $022 + 122^{-}$ (dotted), and $022 + 122^{+}$ (dash-dotted). 
All posteriors are largely consistent with one another. 
The 90\% credible upper limit on $\ell$ is 53.6 km using only the 022 mode, and 62.1 km, 62.3 km, and 62.9 km when including the $122^{0}$, $122^{-}$, and $122^{+}$ modes, respectively. 
Clearly, the constraints weaken when introducing higher modes in the fitting, because these additional modes do not contribute enough information to the likelihood\footnote{This is because we are starting the quasi-normal fitting in the quasi--linear regime, about $10 M_z$ after the peak of the amplitude, where higher overtones are subdominant to the dominant mode and do not contribute much signal-to-noise ratio.}, while they do enlarge the prior volume. In fact, the signal evidence obtained using the waveform model that contains both the $nlm = 022$ and $122^{0}$ modes is about $\exp(2.29) \approx 10$ times smaller than that using just the 022 mode. Nonetheless, the 90\% confidence limits are still very similar to each other, and the overall shape of the posteriors is also similar.
These results demonstrate that the inclusion of an overtone does not have a significant impact on the inferred constraints, and, in particular, they do not significantly alter the posterior.

\subsection{Comparison with ringdown-alone tests of cubic gravity}

This work is the first to accurately and robustly search for curvature-coupled scalar fields from ringdown signals alone. 
One other work has considered a somewhat similar ringdown test, that of cubic gravity \cite{Maenaut:2024oci}, which is a theory that does not include any additional scalar fields. 
The Lagrangian density of cubic gravity theory consists of an additional term $\ell_{c}{}^4 \mathscr{R}^3$, where $\ell_c$ is its coupling length scale and $\mathscr{R}$ is a curvature quantity. 
The widths of the posteriors we found for the curvature-coupled scalar fields we studied are comparable to those found for $\ell_c$ in \cite{Maenaut:2024oci}. 
This numerical coincidence can be understood physically by considering the field with, for example, a dynamical Chern-Simons coupling (though similar reasoning applies to other couplings considered in this work) as follows. 
In terms of the scaled scalar field $\vartheta_2 = \ell_{2}{}^{-2} \varphi_2$, the Lagrangian density in dynamical Chern-Simons gravity is
\begin{equation}
\begin{split}
16 \pi \mathscr{L} = & R + \ell_2{}^4 \left( \vartheta_2 \mathscr{P} - \frac{1}{2} \nabla_{\mu} \vartheta_2 \nabla^{\mu} \vartheta_2 \right). 
\end{split}
\end{equation}
The scaled scalar field $\vartheta_2$ obeys $\Box \vartheta_2 + \mathscr{P} = 0$ and thus, $\varphi_2 = - \Box^{-1} \mathscr{P}$. 
Heuristically, given a characteristic length scale $\lambda$, the curvature scales as $\lambda^{-2}$, so $\mathscr{P} \sim \lambda^{-4}$ and $\Box^{-1} \sim \lambda^2$, yielding $\vartheta_2 \sim \lambda^{-2}$. 
The dynamical Chern-Simons corrections to the Lagrangian, $\vartheta_2 \mathscr{P}$ and $\nabla_{\mu} \vartheta_2 \nabla^{\mu} \vartheta_2 $, then scale as $\lambda^{-6}$, roughly cubic in curvature scale, as is also the case in cubic gravity theories. 
This explains why our constraints on quadratic gravity are similar to those on cubic theories. 

\end{document}